\documentclass[11pt,a4paper]{article} 
\pdfoutput=1
\usepackage{jcappub}
\usepackage{enumerate}
\usepackage{epsfig}
\usepackage{wasysym} 
\usepackage{mathrsfs} 
\usepackage{graphicx} 
\usepackage{amsfonts} 
\usepackage{amsbsy} 
\usepackage{amscd}
\usepackage{slashed} 
\usepackage{multirow} 
\usepackage{gensymb}
\usepackage{subcaption}
\usepackage{natbib}
\allowdisplaybreaks
\usepackage[section]{placeins}
\usepackage{color}
\definecolor{myred}{rgb}{0.6,0,0} 
\definecolor{myblue}{rgb}{0,0.2,0.4}
\definecolor{mygreen}{rgb}{0,0.9,0.1}
\definecolor{hc}{rgb}{.9,0.1,0.7}
\definecolor{hcout}{rgb}{.9,0.7,0.9}
\definecolor{Orange}{rgb}{1.,0.65,0.}

\makeatletter

\newcommand{\fmslash}[2][0mu]{%
  \mathchoice
    {\fmsl@sh\displaystyle{#1}{#2}}%
    {\fmsl@sh\textstyle{#1}{#2}}%
    {\fmsl@sh\scriptstyle{#1}{#2}}%
    {\fmsl@sh\scriptscriptstyle{#1}{#2}}}
\newcommand{\fmsl@sh}[3]{%
  \m@th\ooalign{$\hfil#1\mkern#2/\hfil$\crcr$#1#3$}}
\makeatother

\newcommand{\lsim}{{\;\raise0.3ex\hbox{$<$\kern-0.75em\raise-1.1ex\hbox{$\sim$}}\;}}
\newcommand{\gsim}{{\;\raise0.3ex\hbox{$>$\kern-0.75em\raise-1.1ex\hbox{$\sim$}}\;}}

\usepackage{array}
\newcolumntype{C}[1]{>{\centering\arraybackslash$}p{#1}<{$}}

\usepackage{bm} 
\newcommand{\be}{\begin{equation}}
\newcommand{\ee}{\end{equation}}
\newcommand{\bes}{\begin{equation*}}
\newcommand{\ees}{\end{equation*}}
\newcommand{\bea}{\begin{eqnarray}}
\newcommand{\eea}{\end{eqnarray}}
\newcommand{\beas}{\begin{eqnarray*}}
\newcommand{\eeas}{\end{eqnarray*}}
\title{Search for decaying heavy dark matter in an effective interaction framework: a comparison of $\gamma$-ray and radio observations}
\author[a]{Avirup Ghosh,} \author[a]{Arpan Kar,} \author[b]{Biswarup 
Mukhopadhyaya}
\affiliation[a]{Regional Centre for Accelerator-based Particle
  Physics, Harish-Chandra Research Institute, HBNI, Chhatnag Road,
  Jhunsi, Allahabad - 211 019, India}
\affiliation[b]{Department of Physical Sciences, Indian Institute of Science Education and Research, Kolkata, Mohanpur - 741246, India.}  
  
\emailAdd{avirupghosh@hri.res.in}
\emailAdd{arpankar@hri.res.in} 
\emailAdd{biswarup@iiserkol.ac.in}

\abstract{We investigate and compare the possibilities of observing decaying dark matter (DM) in 
$\gamma$-ray and radio telescopes. The special emphasis  of the study is on a scalar heavy DM particle with mass in the trans-TeV range. DM decays, 
consistent with existing limits on the lifetime, are assumed to be driven by higher dimensional effective operators. 
We consider both two-body decays of a scalar dark particle and a dark sector having three-body decays, producing two standard model particles. 
It is found that the Fermi-LAT data on isotropic $\gamma$-ray background provides the best constraints so far, although the CTA telescope may be more effective 
for decays where one or two photons are directly produced. In all cases, deeper probes of the effective operators are possible in the upcoming SKA radio telescope 
with a few hundred hours of observation, using the radio synchrotron flux coming from energetic electrons produced in the decay cascades within dwarf spheroidal 
galaxies. Finally, we estimate how the SKA can constrain the parameter space spanned by the galactic magnetic field and the diffusion coefficient, 
if observations consistent with $\gamma$-ray data actually take place.}

\subheader{HRI-RECAPP-2020-002}
\date{\today}


\begin{document}
\maketitle
\newpage

\section{Introduction}
\label{sec:intro}
Dark matter (DM) has been emerging as an unavoidably large component of the
energy density of our universe. Perceiving it as resulting from some
hitherto unidentified invisible elementary particle has gained ground as
an acceptable explanation. The mass of a cold dark matter particle
is rather difficult to ascertain from existing evidence. A particularly
challenging possibility is that of a DM particle in the mass range exceeding
a TeV. Such an invisible particle, even if weakly interacting, is largely
unconstrained from direct search experiments \cite{Aprile:2017iyp,PhysRevLett.118.251302}. 
On the other hand, one does not expect such a heavy DM to lead to perceptible missing-$E_T$  signals at, say, the large hadron collider (LHC).
In scenarios such as supersymmetry, where the production of coloured new particles can lead via cascades to DM production, 
the search limit on DM mass is unlikely to go up appreciably above a TeV. This limit is even less in cases where one has to depend 
on Drell-Yan processes for DM production~\cite{SHCHUTSKA2016656,ATL-PHYS-PUB-2014-010,CMS-PAS-FTR-13-014}.
One may therefore have to depend on indirect signals of trans-TeV DM particles. This makes it imperative to think of as many
independent indirect signals as possible.

While independent signals of a stable DM particle come mostly from its
annihilation into standard model (SM) particles \cite{Cuoco:2017iax,TheFermi-LAT:2017vmf,Fermi-LAT:2016uux,Regis:2017oet,Natarajan:2013dsa}, it is not inconceivable
that such a particle is not fully stable, though its lifetime must exceed
the age of the universe by at least ten orders of magnitude~\cite{Blanco:2018esa,Lu:2015pta}. The exact limit on the lifetime
is decided by the dominant decay mode, such as particle-antiparticle pairs,
photons, or even invisible particles like neutrinos, as also on the spin
of the DM particle \cite{Blanco:2018esa,Ando:2015qda,Essig:2013goa,Regis:2017oet,Cohen:2016uyg}. Observations from Fermi-LAT~\cite{Ackermann:2014usa}, 
AMS-02~\cite{Aguilar:2015ooa,Aguilar:2014mma}, HESS~\cite{Abramowski:2014vox,Abramowski:2014tra}, IceCube~\cite{Aartsen:2018mxl} etc. 
contribute to the existing limits. 

An upcoming telescope array that can improve our understanding of
trans-TeV DM is the Cherenkov telescope Array (CTA)~\cite{Consortium:2010bc,Morselli:2017ree} which looks for
highly energetic gamma-rays, arising either from direct decay
of quasi-stable DM or from cascades. Considerable attention has been already paid
to the potential of CTA observation in photonic annihilation of DM pairs~\cite{Silverwood:2014yza,Lefranc:2015pza,Lefranc:2016dgx,Hiroshima:2019wvj}.
It is comparably interesting to investigate similar potential of the CTA as well  as the
already existing Fermi-LAT, as far as decaying DM is concerned~\cite{Viana:2019ucn,Blanco:2018esa,Pierre:2014tra}. 

Our aim here is to study decay of a heavy DM particle by parameterizing the decay Lagrangian 
in terms of effective operators, using as illustration scalar DM particle(s) in single-component
as well as multicomponent scenarios.
The importance of multi-messenger data from extra-terrestrial observations in exploring heavy
dark matter decays has been emphasized in the literature \cite{Ishiwata:2019aet}. While earlier work largely stresses, 
for example, gamma-ray and cosmic-ray data in this context,
we investigate here how gamma-ray data are likely to fare in comparison with the radio synchrotron 
fluxes from dwarf spheroidal galaxies (dSph)  arising from high-mass decaying dark matter particles. 
 While the existing radio data provide
some constraints on the DM parameter space \cite{Regis:2017oet}, the picture is likely to improve
considerably when the Square kilometer Array (SKA) telescope starts its operation.
The usefulness of the SKA, especially for high-mass DM, consists not only in the fact
that relatively high frequency ($\gsim$ 400 MHz) radio observation is possible,
but also in the efficacy of separating foregrounds, thanks to its inter-continental
baseline length \cite{SKA}. The prospects of thus exploring trans-TeV stable DM via its
pair-annihilation have already been discussed in recent studies \cite{Kar:2019cqo,Kar:2019mcq}. The present
work is aimed at extending this to decaying heavy DM, and also comparing the
predicted results to existing and future  gamma-ray observations.

The limits on DM decay from $\gamma$-ray data to date come largely from the
isotropic background caused by the intra-galactic DM distribution as
well as the extra-galactic continuum. It has been pointed out that
localised sources do not offer much of an improvement on this in general \cite{Baring:2015sza}, since
the emitted flux from DM decay goes as $\rho_{DM}/m_{DM}$,
as against $\rho^2_{DM}/m^2_{DM}$ in the case of
annihilation~\cite{Ibarra:2013cra}. However, the suppression caused by trans-TeV $m_{DM}$ 
can sometimes be offset by the higher $\rho_{DM}$ in localised sources.
In addition, a ground-based experiment like the CTA has to overcome
backgrounds resulting from the interaction of cosmic-ray (CR) electrons 
and protons with Earth's atmosphere, where a dense source can be helpful to aim at \cite{Garny:2010eg}.
Therefore, from the standpoint of high-mass DM candidates, it is desirable to ultimately
gear the CTA towards observation of   sources that may reveal signals for
optimal values of $\rho_{DM}/m_{DM}$, even
if the isotropic  gamma-ray measurements turn out to have better prospects at present \cite{Viana:2019ucn}.

For radio synchrotron fluxes, on the other hand, one has to depend exclusively
on specific sources with large mass-to-light ratios. A dSph is a popular hunting
ground in this respect. Their low star formation rates also minimize the
astrophysical background \cite{Strigari:2008ib,Strigari:2007at,Strigari:2006rd,Mateo:1998wg}.  The DM decay cascades there,
just as in the case of annihilation, lead to energetic electron-positron pairs
that execute cycloidal motion under the influence of galactic magnetic fields,
leading to radio synchrotron emission whose flux is determined by solving
the appropriate transport equation. 
The `source function' entering into
the transport equation again depends on  $\rho_{DM}/m_{DM}$ for
decaying DM, as opposed to  $\rho^2_{DM}/m^2_{DM}$ in the case of
annihilating pairs. This causes enhanced fluxes for a dSph with high DM
density profile, when one is looking at the decays of trans-TeV DM.

Keeping the above observations in mind, we focus here on gamma-ray
predictions and constraints vis-a-vis those for SKA, mostly using the dSph Draco as
example.  As has been mentioned above, scalar DM particles have been
used to illustrate our point, although the conclusions are easily
extendable to a fermionic dark sector. We consider various $SU(2)_L
\times U(1)_Y$ invariant effective operators, as listed in the next
section, driving DM decays in various channels. We have included
two-body decays of the DM, as also three-body decays of one quasi-stable particle in the
dark sector decaying into another along with a pair of SM particles.
Decays of the latter generate electron-positron pairs that are the ultimate sources of radio
synchrotron emission.
For each case, we compare the upper limit on the decay width
from gamma-ray data  with those
expected from the SKA. These can be translated into limits on the coefficients of the effective operators, including the Wilson coefficients and the
suppression scale of the operators. 
As we shall see, the upper limits mostly come from the Fermi-LAT
data on isotropic gamma-ray, whenever they are available. The projected CTA sensitivity
in such cases mostly require decay widths that are already ruled out \cite{Viana:2019ucn}. Therefore,
for such cases we compare the potential of radio signal measurements at SKA with
Fermi-LAT observations on isotropic gamma-rays,
find that the former can probe deeper into DM parameter space. 

An exception is the situation where  the DM mass exceeds about 1 TeV, and  the
DM decays directly into one or two photons. The available Fermi-LAT data in such a case
offer no limits \cite{Garny:2010eg}. This is where the projected CTA measurements have been compared here
with the corresponding expectations from the SKA. While the SKA predictions
pertain to an illustrative dSph, namely Draco, the CTA projections  shown here
still focus on the isotropic gamma-ray observation, since for the common dSph's
are found to be less promising for CTA in terms of DM decays \cite{Viana:2019ucn}.  We try to
understand how the predicted signals (or their absence) at
the SKA can yield information on the space of astrophysical parameters
in the dSph, spanned by quantities like the galactic magnetic field
and the diffusion coefficient.

The effective operators listed by us are assumed to be responsible for DM decays in
galaxies. However, we take a model-independent  view of the 
relic density \cite{Ade:2015xua}, by not ruling 
out other production/annihilation channels.  The assumption inbuilt in the present study is that
only the effective operators under consideration here are responsible for indirect
DM decay signals. 

The paper is organized as follows: In sec.~\ref{sec:effop} we have parametrized the DM decay
into gauge boson as well as fermion pairs in terms of higher dimensional operators. 
Sec.~\ref{sec:signals} contains a brief discussion of the astrophysical signals of decaying DM,
namely the $\gamma$-ray flux as well as the radio synchrotron flux. We have presented our findings in
sec.~\ref{sec:limits}. Finally we conclude and summarize in sec.~\ref{sec:Conclusion}. The 
necessary formulae used for our analysis can be found in Appendices~\ref{AppendixA} and \ref{AppendixB}.

\section{Effective operators}
\label{sec:effop}

The standard model(SM) of Particle Physics does not contain any suitable
DM candidate. Thus the extension of the SM particle content is inevitable. 
DM and its stability are frequently explained by the postulation
of one or more new particles and some new symmetry, whose most popular (but by no means unique) formulation is the discrete group $Z_2$. 
With such a discrete symmetry the part of the particle spectrum which is odd under
it constitutes a `dark sector'. There can be 
decays within the dark sector, till the lightest particle in that sector
is reached, the latter becoming stable and contributing to the relic density \cite{Biswas:2017ait,Essig:2013goa,Garny:2012vt}.
Alternatively, one may have no such symmetry surviving at the mass scale of the
DM particle, and allow the latter to decay, albeit very slowly \cite{Arcadi:2014tsa,Arcadi:2013aba}.
The basic requirement for such decays is that the lifetime should exceed the age of the universe. However, the constraint is most stringent if the final state consists of 
visible particles, due to limits from, for example, cosmic-ray photons as well as positrons and antiprotons \cite{Ibarra:2013cra}.

Parametrization of the decay of a DM candidate 
by dimension-5 effective operators is strongly constrained~\cite{Slatyer:2017sev},
since in that case
\begin{equation}
\tau_{DM}\simeq\,6.58\times10^4\,{\rm s}\left(\frac{m_{DM}}{1\,{\rm TeV}}\right)^{-3}\left(\frac{\Lambda}{10^{19} {\rm GeV}}\right)^{2},
\end{equation} 
leaving out factors dependent on the spin of the DM particle. This
exceeds the requisite lower limit only when $m_{DM}\leq\!\mathcal{O}({\rm MeV})$, even with 
$\Lambda\simeq\!10^{19}$GeV \cite{Canetti:2012kh}. For most of the DM parameter space, one
thus finds it more consistent to parametrize all the decay interactions of 
the DM by dimension-6 operators, the suppressant scale $\Lambda$ being the 
mass scale of the new physics responsible for generating such interactions. 

As has been mentioned in the introduction, we simplify our analysis
by confining ourselves to a scalar dark sector, though the features related
to its detection pointed out by us apply to particles with spin as well.
We consider two possible scenarios within this category:
\begin{enumerate}
\item A single-component dark matter which is quasi-stable
over the age of the universe and has two-body decays into SM particles. 
\item A multicomponent (two-component) scenario where the heavier of the two dark sector
members is quasi-stable and decays into the lighter, stable one,  
along with  visible SM particles. 
\end{enumerate}

  We outline these two scenarios below 
  \footnote{ In principle, both of these features
  may be found in a multicomponent dark sector where the lightest 
  particle, too, is long-lived but unstable. The analysis of such a 
  scenario requires multiple effective interactions to be operative
  at the same time. We simplify our analysis by taking one type of effective operator at a time, 
  where the nature of effective interactions gets related more transparently 
  to aspects of DM decay observations in the $\gamma$-ray and radio ranges.}.

\subsection{Single-component scalar dark matter}
\label{sec:singlecompDM}

Following the above observation, we postulate dimension-6 terms as being responsible for
DM decays. Modulo some hitherto unspecified symmetry, broken by the vacuum expectation value (vev) of scalar DM field $\phi$, 
the dimension-6 operators reduced to dimension-5 ones, dictating two-body DM decays
\footnote{Smallness of the effective dimension-5 operators can be justified by an appropriate vev for $\phi$. Similarly decays like
$\phi$ to a pair of SM higgs is assumed here to be negligible, by postulating a near-vanishing interaction between the dark sector scalars and the SM higgs.}. 
The corresponding dimension-5 operators can be parameterised as \cite{Mambrini:2015sia,Grzadkowski:2010es}: 
\begin{eqnarray}
-\mathcal{L}_{dim-5}\,&\supset\,&\,-\mathcal{L}^{\rm gauge}_{dim-5}-\mathcal{L}^{\rm fermion,1}_{dim-5}-\mathcal{L}^{\rm fermion,2}_{dim-5}
\end{eqnarray}  
where
\begin{eqnarray}
-\mathcal{L}^{\rm gauge}_{dim-5}&=&\frac{f_{WW}}{\Lambda}\phi\,W^{a\,\mu\,\nu}W^{a}_{\mu\,\nu}+\frac{f_{BB}}{\Lambda}\phi\,B^{\mu\,\nu}B_{\mu\,\nu},\nonumber\\
-\mathcal{L}^{\rm fermion,1}_{dim-5}&=&\phi\,\bigg(\frac{f_{QQ}}{\Lambda}\bar{Q}_{L}\gamma^{\mu}D_{\mu}Q_{L}+\frac{f_{uu}}{\Lambda}\bar{u}_{R}\gamma^{\mu}D_{\mu}u_{R}+\frac{f_{dd}}{\Lambda}\bar{d}_{R}\gamma^{\mu}D_{\mu}d_{R}\bigg)\nonumber\\
&&\hspace{0.8cm} +\phi\,\bigg(\frac{f_{l_L l_L}}{\Lambda}\bar{l}_{L}\gamma^{\mu}D_{\mu}l_{L}+\frac{f_{l_R l_R}}{\Lambda}\bar{l}_{R}\gamma^{\mu}D_{\mu}l_{R}\bigg),\nonumber\\
&=& \phi\,\frac{f_{qq}}{\Lambda}\bigg(\bar{Q}_{L}\gamma^{\mu}D_{\mu}Q_{L}+\bar{u}_{R}\gamma^{\mu}D_{\mu}u_{R}+\bar{d}_{R}\gamma^{\mu}D_{\mu}d_{R}\bigg)\nonumber\\
&&\hspace{0.8cm} +\phi\,\frac{f_{ll}}{\Lambda}\bigg(\bar{l}_{L}\gamma^{\mu}D_{\mu}l_{L}+\bar{l}_{R}\gamma^{\mu}D_{\mu}l_{R}\bigg),\nonumber\\
-\mathcal{L}^{\rm fermion,2}_{dim-5}&=&\phi\bigg[\frac{f_{uuH}}{\Lambda}\bar{Q}_{L}u_{R}\tilde{H}+\frac{f_{ddH}}{\Lambda}\bar{Q}_{L}d_{R}H+\frac{f_{llH}}{\Lambda}\bar{l}_{L}l_{R}H+h.c\bigg],
\label{eqn:dim5_single}
\end{eqnarray} 
with $\Lambda$ being the suppression scale. Here $\phi$ is $SU(2)_L\times U(1)_Y$-singlet thus making each operator invariant under the electroweak group. 
While presenting our results, we will however consider only one operator to be dominant at a time, 
for the sake of simplicity. 
In each such case the two-body decays in the respective final states is taken to dominate DM decay, the three-body
decays driven by the corresponding operators being understandably suppressed.   
For simplicity we have also assumed that $f_{QQ}=f_{uu}=f_{dd}=f_{qq}$ and 
$f_{l_L l_L}=f_{l_R l_R}=f_{ll}$ while presenting our results. Expressions for the two-body partial decay widths are
given in Appendix~\ref{AppendixA}.

\subsection{Multicomponent scalar dark sector}
\label{sec:multicompDM}
As an alternative scenario, we consider a multicomponent dark sector containing two SM singlet
$Z_{2}$-odd real scalars $\phi_{2}$ and $\phi_{1}$. We assume
$\phi_{2}$ (identifying $M_{2}\equiv\!m_{DM}$, as the mass of the 
decaying dark matter) is heavier than $\phi_{1}$ (with mass $M_{1}$) 
and $\phi_{2}$ decays to $\phi_{1}$ \cite{Ghosh:2019jzu}. We parametrize these 
decay modes in terms of several dimension-6 operators \cite{Grzadkowski:2010es}
\footnote{We have neglected the dimension-4 interaction term $\lambda_{12} \phi_2 \phi_1 H^{\dagger}H$ 
compared to the dimension-6 effective operators presented in Eqn.~\ref{eqn:dim6_multi}.},
\begin{eqnarray}
-\mathcal{L}_{dim-6}\,&\supset\,&\,-\mathcal{L}^{\rm gauge}_{dim-6}-\mathcal{L}^{\rm fermion,1}_{dim-6}-\mathcal{L}^{\rm fermion,2}_{dim-6}
\end{eqnarray}  
where
\begin{eqnarray}
-\mathcal{L}^{\rm gauge}_{dim-6}&=&\frac{f_{WW}}{\Lambda^{2}}\phi_{2}\phi_{1}W^{a\,\mu\,\nu}W^{a}_{\mu\,\nu}+\frac{f_{BB}}{\Lambda^{2}}\phi_{2}\phi_{1}B^{\mu\,\nu}B_{\mu\,\nu}+\frac{f_{B}}{\Lambda^{2}}\left(\partial_{\mu}\phi_{2}\partial_{\nu}\phi_{1}-\partial_{\nu}\phi_{2}\partial_{\mu}\phi_{1}\right)B^{\mu\,\nu},\nonumber\\
-\mathcal{L}^{\rm fermion,1}_{dim-6}&=&\phi_{2}\overset{\leftrightarrow}{\partial}_{\mu}\phi_{1}\bigg(\frac{f_{QQ}}{\Lambda^{2}}\bar{Q}_{L}\gamma^{\mu}Q_{L}+\frac{f_{uu}}{\Lambda^{2}}\bar{u}_{R}\gamma^{\mu}u_{R}+\frac{f_{dd}}{\Lambda^{2}}\bar{d}_{R}\gamma^{\mu}d_{R}\nonumber\\
&&\hspace{1.5cm} +\frac{f_{l_L l_L}}{\Lambda^{2}}\bar{l}_{L}\gamma^{\mu}l_{L}+\frac{f_{l_R l_R}}{\Lambda^{2}}\bar{l}_{R}\gamma^{\mu}l_{R}\bigg),\nonumber\\
&=&\phi_{2}\overset{\leftrightarrow}{\partial}_{\mu}\phi_{1}\frac{f_{qq}}{\Lambda^{2}}\bigg(\bar{Q}_{L}\gamma^{\mu}Q_{L}+\bar{u}_{R}\gamma^{\mu}u_{R}+\bar{d}_{R}\gamma^{\mu}d_{R}\bigg)\nonumber\\
&&\hspace{1.5cm} +\phi_{2}\overset{\leftrightarrow}{\partial}_{\mu}\phi_{1}\frac{f_{ll}}{\Lambda^{2}}\bigg(\bar{l}_{L}\gamma^{\mu}l_{L}+\bar{l}_{R}\gamma^{\mu}l_{R}\bigg),\nonumber\\
-\mathcal{L}^{\rm fermion,2}_{dim-6}&=&\phi_{2}\phi_{1}\left[\frac{f_{uuH}}{\Lambda^{2}}\bar{Q}_{L}u_{R}\tilde{H}+\frac{f_{ddH}}{\Lambda^{2}}\bar{Q}_{L}d_{R}H+\frac{f_{llH}}{\Lambda^{2}}\bar{l}_{L}l_{R}H+h.c\right].
\label{eqn:dim6_multi}
\end{eqnarray}

Unlike the case of sec.~\ref{sec:singlecompDM}, the energy distribution
of the primary decay products in this case depends on the lorentz
structure of the matrix element itself (see Appendix~\ref{AppendixB}).
We have considered three-body decays~\footnote{The four-body decays 
e.g. $\phi_2 \rightarrow \phi_1 h f \bar{f}, \phi_1 W^{+}W^{-}Z$ 
are sub-dominant compared to the three-body decays driven by the same operators due to phase space suppression and hence have been neglected in our analysis.}
into bosonic final states $\phi_2\rightarrow\phi_1W^+W^-,\phi_1ZZ,\phi_1Z\gamma,
\phi_1\gamma\gamma$, $\phi_1\!Z$~\footnote{Note that $\phi_2\rightarrow\phi_1\!\gamma$ is suppressed from angular momentum conservation.} 
as well as fermionic final states $\phi_2\rightarrow\phi_1b\bar{b},
\phi_1t\bar{t},\phi_1\tau^+\tau^-$. 
As in the case of single-component dark matter, here also we have  assumed that $f_{QQ}=f_{uu}=f_{dd}=f_{qq}$ and $f_{l_L l_L}=f_{l_R l_R}=f_{ll}$, for simplicity.
In addition to the DM mass $M_2$ and the Wilson coefficient driving the decay under consideration,
$\Delta M = M_2 - M_1$ is also a parameter that affects 
the $\gamma$-ray and radio signals.  
While determining the signals of $\phi_2$ decay from various astrophysical objects
we have assumed $\phi_2$ density to be same as the DM density of that object 
{\it i.e.} $\rho_{\phi_2}=\rho_{\rm DM}$. For $\rho_{\phi_2} < \rho_{\rm DM}$ the 
results presented in sec.~\ref{sec:limits} get relaxed by a factor of 
$\rho_{\phi_2}/\rho_{\rm DM}$. 

\section{Astrophysical signals of decaying DM}
\label{sec:signals}
The SM particles produced in DM decay lead to further cascades. Charged particles in such cascades
can produce $e^{\pm}$ which subsequently emit radio synchrotron signal as a result of cycloidal motion under galactic magnetic field. 
Such radio signals can be detected by the SKA radio telescope if the frequency lies in the range 50 MHz - 50 GHz \cite{SKA}. 
The high frequency range should serve as especially important signal of highly energetic electrons produced from the decay of trans-TeV DM.
On the other hand, the source of energetic $\gamma$-rays is usually neutral pions produced in cascades. 
In addition, directly produced photons can also contribute to the $\gamma$-ray as well as radio signals, when the effective operators couple $\phi/\phi_{1,2}$ directly to electroweak field strength tensors.

\subsection{DM induced $\gamma$-ray flux}
\label{sec:gammasignals}

The differential $\gamma$-ray flux originating from the SM final states
(e.g. $W^+W^-,ZZ,Z\gamma,\gamma\gamma,\bar{b}b$, $\bar{t}t,\tau^+\tau^-$) of
DM decay inside our galaxy is given by \cite{Ibarra:2013cra}
\begin{equation}
\frac{d\Phi_{\rm Gal}}{dE_\gamma} (E_\gamma,\Omega) = \frac{\Gamma}{4 \pi m_{DM}} \times \sum_f \frac{dN^{\gamma}_f}{dE_\gamma}(E_\gamma) B_f \times J_{d}(\Omega)
\label{eqn:siggamflux_Gal}
\end{equation}
where $\Gamma$ is the total decay width (into all allowed channels)
of the DM particle, $m_{DM}$ is the mass of the decaying dark matter 
particle, $dN^{\gamma}_f/dE_\gamma$ is the differential distribution of the 
$\gamma$-ray photons produced per decay for the final state $f$ with branching ratio $B_f$. This differential distribution is calculated using 
\cite{micrOMEGAs,Belanger:2010gh}.
The astrophysical factor contributing to the determination of this
flux is encoded in the J-factor, $J_d$ for decaying DM which can be 
expressed as
\begin{equation}
J_{d} (\Omega) = \int_{l.o.s} ds \hspace{1mm} \rho_d(r(s,\Omega))
\label{eqn:Jfactor}
\end{equation}
where $\rho_d(r)$ is the density of decaying DM inside our galaxy. 
We have considered $\rho_d(r)$ to be a standard Navarro-Frenk-White (NFW) \cite{Navarro:1995iw} profile:
\begin{equation}
\rho_d (r)=\frac{\rho_0}{\left(r/r_s\right)\left(1+r/r_s\right)^2}
\label{NFW}
\end{equation}
where $r_s=20\,$kpc and $\rho_0$ is such that the local DM density $\rho_d(r=8.25 {\rm kpc})= 0.4$ GeV/cm$^3$ \cite{2009ApJ...704.1704B,Gillessen:2008qv}.

On the other hand, DM decay inside of our galaxy also gives rise to $e^{\pm}$ which can transfer their energy to photons of CMB,
dust scattered light and starlight via Inverse Compton Scatterings(ICS). The energy distribution of ICS gamma-rays are given by~\cite{Cirelli:2010xx}:
\begin{equation}
\frac{d\Phi_{\rm ICS}}{dE_\gamma}(E_\gamma, \Omega) =\frac{1}{4\pi E_\gamma} \int_{l.o.s} ds \times 2\int_{m_e}^{m_{\rm DM}/2} dE_e \mathcal{P}_{\rm ICS}(E_\gamma,E_e,\vec{r}) \frac{dn_{e}}{dE_e}(E_e,\vec{r})
\label{eqn:ICSflux}
\end{equation}   
where $\mathcal{P}_{\rm ICS}$ is the ICS power spectrum in Klein-Nishina regime which includes the distribution of photons in the 
inter-steller radiation field of CMB, dust scattered light and starlight~\cite{Porter:2005qx}. 
On the other hand, $\dfrac{dn_{e}}{dE_e}(E_e,\vec{r})$ is the steady-state $e^{\pm}$ distribution obtained from the diffusion-loss equation:
\begin{equation}
\nabla \left[D(E_e,\vec{r})\nabla \left( \frac{dn_{e}}{dE_e}(E_e,\vec{r})\right)\right]+\frac{\partial}{\partial E}\left(b(E_e,\vec{r}) \frac{dn_{e}}{dE_e}(E_e,\vec{r})\right)+Q(E_e,\vec{r})=0
\label{eqn:transporteqICS}
\end{equation}
where $D(E_e,\vec{r})$ is the diffusion parameter which have been taken to be position independent for simplicity and we have used 
$D(E_e)=3.33\times 10^{27}\,{\rm cm}^2\,{\rm s}^{-1} \left(E_e/{\rm GeV}\right)^{0.7}$~\cite{Delahaye:2007fr,Donato:2003xg}. 
The eqn~\ref{eqn:transporteqICS} have been solved in a cylindrical diffusion zone of height of 8 kpc and radius 20 kpc~\cite{Cirelli:2010xx}. 
For the energy-loss term  $b(E_e,\vec{r})$ we have used the parametrization provided in~\cite{Cirelli:2010xx}. 
The third term of eqn~\ref{eqn:transporteqICS} is the source term:
\begin{equation}
Q(E_e,\vec{r})= \frac{\rho_d(\vec{r}) \times \Gamma}{m_{\rm DM}}\underset{f}{\sum}\frac{dN^f_e}{dE_e}(E_e)B_f 
\end{equation}   
where $\rho_d(\vec{r})$ is given in eqn~\ref{NFW} and $\dfrac{dN^f_e}{dE_e}(E_e)$ is the $e^{\pm}$ differential distribution 
produced per DM decay in the final state $f$. 

As for the galactic contribution to $\gamma$-rays from
DM decay, the direction of observation ($\Omega$) in eqns.\ref{eqn:siggamflux_Gal} and~\ref{eqn:ICSflux} 
has been defined for the Fermi-LAT observation region $|b| > 20^{\degree}, |l|< 180^{\degree}$ \cite{Ackermann:2014usa,Blanco:2018esa}, 
for which the astrophysical sources of isotropic gamma-rays are well resolved.

The gamma-rays originating from the DM decay outside of our galaxy also contributes to Fermi's measurement of IGRB. The extra-galactic 
contribution to the gamma-ray flux is \cite{Ibarra:2013cra},
\begin{equation}
\frac{d\Phi_{\rm EG}}{dE_\gamma}(E_\gamma)=\frac{ \Gamma}{4\pi m_{\rm DM}} c\,\Omega_{\rm DM}\rho_c\int^{\infty}_0 \frac{dz}{H(z)}\sum_f \left(\frac{dN^\gamma_f}{dE^{\prime}_\gamma} \left[E_\gamma (1+z)\right]  B_f \right)e^{-\tau(E_\gamma, z)}
\end{equation}
where $\rho_c=4.9\times 10^{-6}$ GeV/cm$^3$, $\Omega_{\rm DM}=0.26$,
$H(z)=H_0 \sqrt{\Omega_\Lambda + \Omega_m (1+z)^3}$ 
with $\Omega_\Lambda=0.69$,$\Omega_m=0.31$ and $e^{-\tau(E_\gamma, z)}$ signifies the attenuation 
due to extra-galactic absorption which has been taken from~\cite{Cirelli:2010xx}. 

Thus the total $\gamma$-ray flux from DM decays,
\begin{equation}
\frac{d\Phi}{dE_\gamma} = \frac{d\Phi_{\rm Gal}}{dE_\gamma} + \frac{d\Phi_{\rm ICS}}{dE_\gamma} + \frac{d\Phi_{\rm EG}}{dE_\gamma}
\label{eqn:Totgammaflux}
\end{equation}
which have been compared with the Fermi's result of IGRB~\cite{Ackermann:2014usa} while deriving the limits presented in sec.~\ref{sec:limits}. Having thus taken all contributions into account, it is found that for DM masses up to $300\,{\rm GeV}$,  
$\Phi_{\rm Gal}+\Phi_{\rm EG}$ largely determines the limit, while $\Phi_{\rm ICS}$ and $\Phi_{\rm EG}$ play decisive roles for even 
higher masses. A caveat to be added here, however, is that ICS may become  dominant if the DM directly decays into $e^{\pm}/\mu^{\pm}$ pairs, 
in contrast to the channels we have considered here. We refer the reader to \cite{Blanco:2018esa} for details. 

\begin{figure*}[ht!]
\centering
  \includegraphics[height=0.4\textwidth, angle=0]{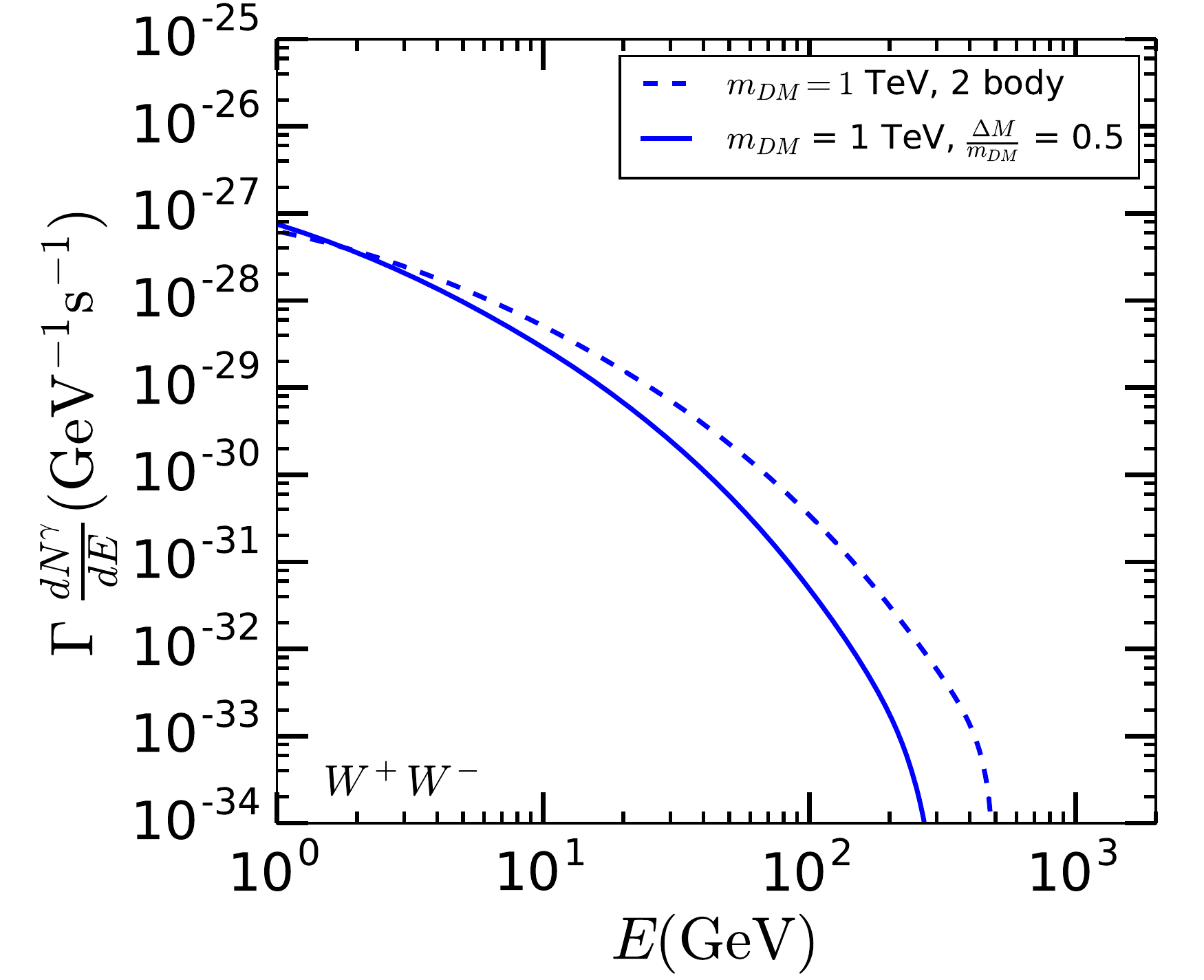}\hspace{2mm}%
  \includegraphics[height=0.4\textwidth, angle=0]{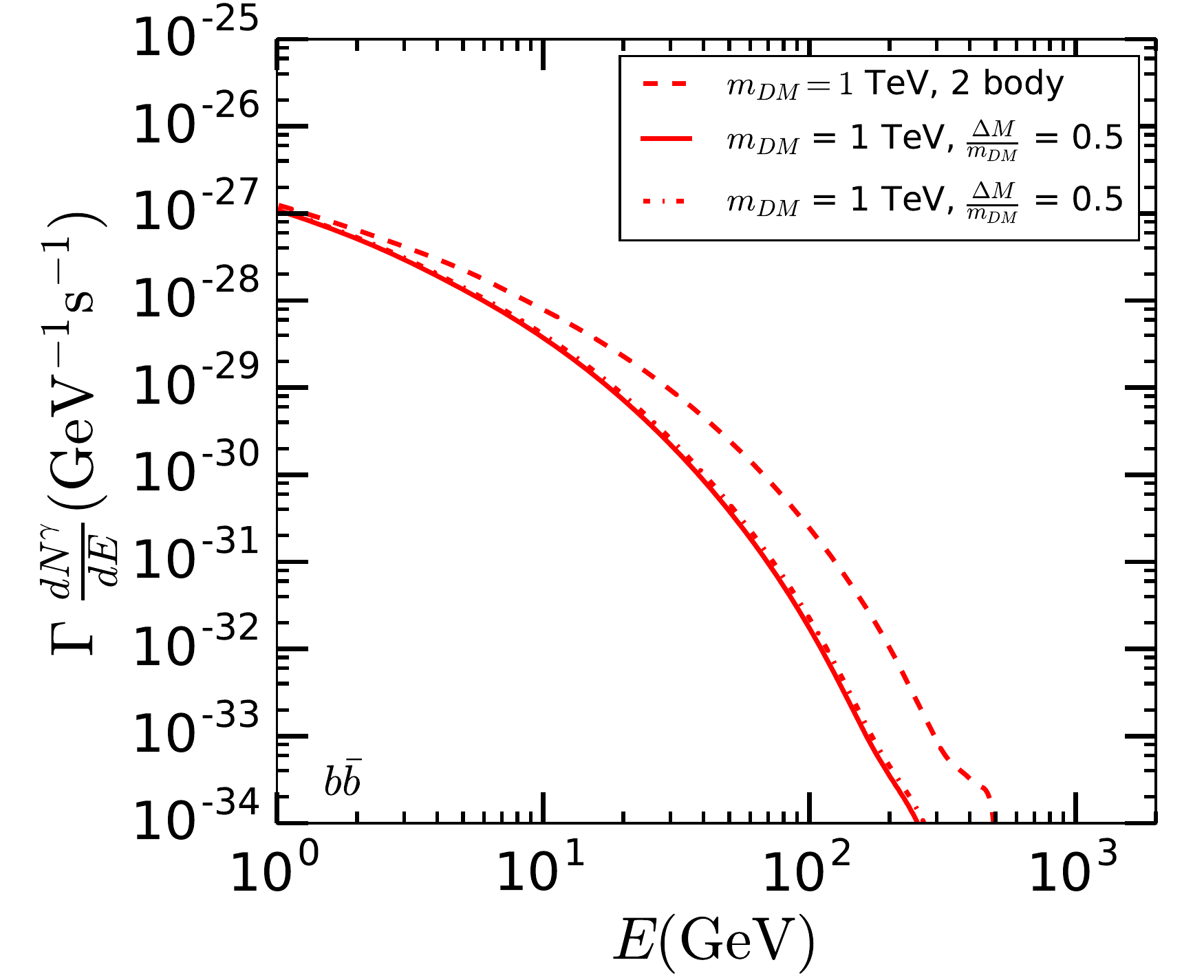}\hspace{2mm}%
 \caption{{\it Left panel:} $\gamma$-ray spectrum ($\Gamma \frac{dN^{\gamma}}{dE}$) for the 
  decay processes $\phi \rightarrow W^+ W^-$ (dashed curve) and $\phi_2 \rightarrow \phi_1 W^+ W^-$ (solid curve). In the latter case, 
  the (normalised) energy distribution 
  of $W^+/W^-$ is governed by Eqn. \ref{eqn:diboson_dist}.
  {\it Right panel:} $\gamma$-ray spectrum for the decay processes $\phi \rightarrow b\bar{b}$ (dashed curve) and 
  $\phi_2 \rightarrow \phi_1 b\bar{b}$ (solid and dashed-dotted curves).
  For three-body decays the (normalised) energy distributions of $b/\bar{b}$ is governed by
  Eqns. \ref{eqn:ffbar_dist1} (solid line) and \ref{eqn:ffbar_dist2} (dashed-dotted line).
  Here $m_{DM} = 1$ TeV ($\frac{\Delta M}{m_{DM}}$ = 0.5 for multicomponent scenario) and $\Gamma = 10^{-28} s^{-1}$.}
  \label{fig:WWgammaflux}
\end{figure*}

The galactic contribution to the total $\gamma$-ray flux in eqn~\ref{eqn:Totgammaflux} has been calculated following \cite{Blanco:2018esa} 
by taking the entire high-latitude sky (i.e. $0^{\degree} < |l| < 180^{\degree}$) for $|b| > 20^{\degree}$ into account. 
Note that this galactic contribution is not truly isotropic and can vary by a factor of 5 within the range of angles considered. 
However, this variation is less than $10\%$ within $45^{\degree}$ of the anti-galactic center (i.e $ 180^{\degree}-45^{\degree} < l < 180^{\degree}+45^{\degree} $). 
Hence, if one takes the contribution towards anti-galactic center only (see~\cite{Cirelli:2012ut,Liu:2016ngs}), 
the galactic $\gamma$-ray flux can be treated more or less isotropic. 
In this case the total $\gamma$-ray flux can decrease at most by a factor of 5 than the flux predicted here and 
thus the limits on the dark matter decay width (presented in 
sec.\ref{sec:limits}) will also be weakened at most by the same factor. 
Since the decay width is proportional to the square of the Wilson coefficients, the corresponding limits on these coefficients 
will be relaxed maximally by a factor of $\sim 2.2$. This will widen the available DM parameter space that can be probed by future radio telescopes like SKA.

We have shown for illustration the $\gamma$-ray distributions ($\Gamma\frac{dN^{\gamma}}{dE_\gamma}$) originating from the 
decays $\phi\rightarrow\,W^+W^-$ ($\phi \rightarrow b\bar{b}$) and 
$\phi_2\rightarrow\phi_1W^+W^-$ ($\phi_2 \rightarrow \phi_1 b\bar{b}$) in the 
left (right) panel of Fig~\ref{fig:WWgammaflux} for a benchmark value of $m_{DM}$,
$\Gamma$ and $\Delta\,M/m_{DM}$. Clearly for the three-body decays a substantial energy is taken away by $\phi_1$, 
thereby softening the corresponding $\gamma$-ray spectrum.

\subsection{DM induced radio flux}
\label{sec:radiosignals}
The SM products of DM decay inside a dSph generate $e^{\pm}$ pairs through 
cascade decays, whose abundance is decided by the source 
function ($Q_e(E,r)$) \cite{Regis:2017oet}:
\begin{equation}
Q_e (E,r) = \Gamma \times \sum_f \frac{dN^{e}_f}{dE}(E) B_f \times \frac{\rho_d (r)}{m_{DM}}.
\end{equation}
where $dN^{e}_f/dE$ is the differential distribution of the 
$e^{\pm}$ produced per decay in the final state $f$ with branching ratio $B_f$.
The differential distribution is obtained using \cite{micrOMEGAs,Belanger:2010gh}.
$\rho_d (r)$ is the DM density profile of a dSph as a function of radial distance $r$ from the centre of the dSph. As already mentioned, 
we have taken the dSph Draco assuming a NFW profile
as given in Eqn. \ref{NFW} with 
$\rho_0 = 1.4$ GeV. cm$^{-3}$ and $r_s = 1.0$ kpc \cite{Colafrancesco:2006he}\footnote{We have checked that, the choice of other profiles such as 
Burkert \cite{Burkert:1995yz,Colafrancesco:2006he} 
or Diemand et al. (2005) \cite{Diemand:2005wv}
(hereafter D05) \cite{Colafrancesco:2006he} keep 
the observed radio flux almost similar.}.
We have used Draco for predicting
the radio signal as various relevant parameters like the J-factor 
are somewhat better constrained for this dSph \cite{Geringer-Sameth:2014yza}. 
However, these parameters are also well-measured for other
dSph's such as Seg1, Carina, Fornax, Sculptor etc \cite{Geringer-Sameth:2014yza,Choquette:2017nqk,Kar:2019hnj}. Draco is used for illustration in our analysis. 

The produced electron(positron) diffuses through the galactic medium and loses energy via several processes like Inverse-Compton 
scatterings(IC), Synchroton radiation(Synch), Coulomb effect, 
bremsstrahlung etc. The final $e^{\pm}$ distribution $\frac{dn_e}{dE}(E,r)$ is obtained by solving the 
differential equation \cite{Beck:2015rna,Colafrancesco:2006he,Colafrancesco:2005ji},  
\begin{equation} 
D(E) \nabla^2 \left(\frac{dn_e}{dE}\right) +
\frac{\partial}{\partial E}\left(b(E) \frac{dn_e}{dE}\right) +
Q_e(E,r) = 0
\label{eqn:transporteq}
\end{equation}
where the diffusion parameter $D(E)$ has been parametrized as $D(E)=D_0\,(\dfrac{E}{\rm GeV})^{0.3}$. The radius of the 
diffusion zone is assumed to be $2.5$ kpc~\cite{Colafrancesco:2006he}. The energy loss coefficient $b(E)$ can be expressed as
\begin{eqnarray} 
b(E) &=& b^0_{IC} \left(\frac{E}{\rm GeV}\right)^2 + b^0_{Synch} \left(\frac{E}{\rm GeV}\right)^2 \left(\frac{B}{\mu G}\right)^2 \nonumber\\
&& + b^0_{Coul} n_e \left[1 + \frac{\log\left(\frac{E/m_e}{n_e}\right)}{75} \right]
+ b^0_{Brem} n_e \left[\log\left(\frac{E/m_e}{n_e}\right) + 0.36 \right],
\end{eqnarray}
where the values of the energy loss coefficients are
$b^0_{IC} \simeq 0.25$, $b^0_{Synch} \simeq 0.0254$, $b^0_{Coul} \simeq 6.13$
$b^0_{Brem} \simeq 1.51$, all in units of $10^{-16}$ GeV s$^{-1}$. $m_e$ denotes the electron mass and $n_e$ is the average thermal electron density
(value of $n_e \approx 10^{-6}$ inside a dSph) \cite{Beck:2015rna,Colafrancesco:2005ji}.

\begin{figure*}[t!]
\centering
  \includegraphics[height=0.4\textwidth, angle=0]{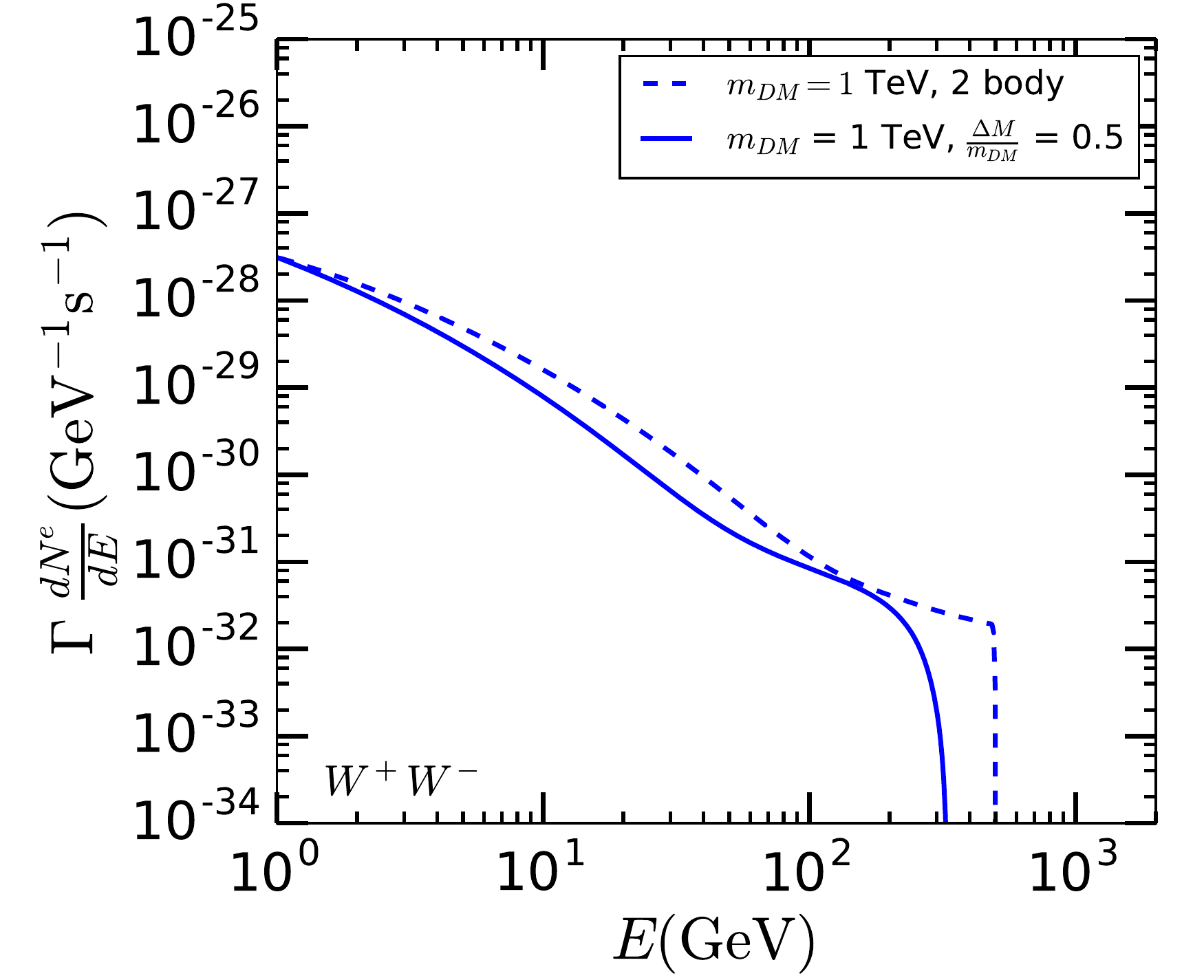}\hspace{2mm}%
  \includegraphics[height=0.4\textwidth, angle=0]{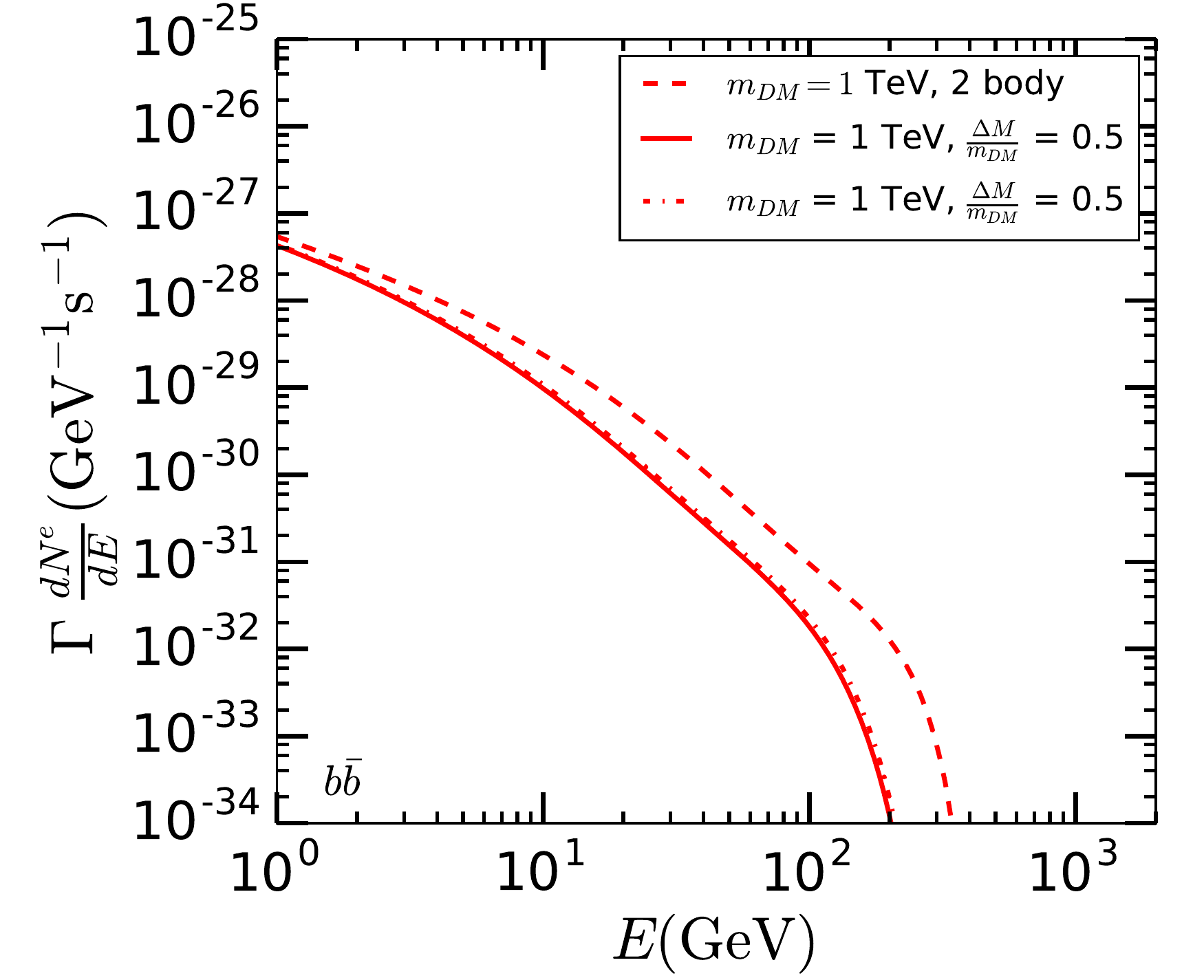}\hspace{2mm}%
  \caption{{\it Left panel:} $e^{\pm}$ spectrum ($\Gamma \frac{dN^{e}}{dE}$) for the 
  decay processes $\phi \rightarrow W^+ W^-$ (dashed curve) and $\phi_2 \rightarrow \phi_1 W^+ W^-$ (solid curve). In the later case, the (normalised) 
  energy distribution of $W^+/W^-$ is governed by Eqn. \ref{eqn:diboson_dist}.
  {\it Right panel:} $e^{\pm}$ spectrum for the decay processes $\phi \rightarrow b\bar{b}$ (dashed curve) and $\phi_2 \rightarrow \phi_1 b\bar{b}$ (solid and dashed-dotted curves).
  For three-body decays the (normalised) energy distributions of $b/\bar{b}$ is governed by
  Eqns. \ref{eqn:ffbar_dist1} (solid line) and \ref{eqn:ffbar_dist2} (dashed-dotted line).
  Here $m_{DM} = 1$ TeV ($\frac{\Delta M}{m_{DM}}$ = 0.5 for multicomponent scenario) and $\Gamma = 10^{-28} s^{-1}$.}
  \label{fig:WWelectronflux}
\end{figure*}

The final radio flux ($S_{\nu}$) as a function of frequency ($\nu$) is 
obtained by folding this $\frac{dn_e}{dE}$ with synchrotron power 
spectrum ($P_{Synch}(\nu,E,B)$) \cite{Kar:2019cqo,Beck:2015rna,Colafrancesco:2006he,Colafrancesco:2005ji} and integrating over the size of the 
emission region of the dSph ($\Delta \Omega$):
\begin{equation}
S_{\nu}(\nu)=\frac{1}{4\pi}\int_{\Delta \Omega}\,d\Omega\int_{l.o.s}\,ds\left(2\overset{m_{DM}/2}{\underset{m_e}{\int}}dE\frac{dn_e}{dE}(r(s,\Omega),E)P_{Synch}(\nu,E,B)\right).
\end{equation}
  
\begin{figure*}[t!]
\centering
  \includegraphics[height=0.4\textwidth, angle=0]{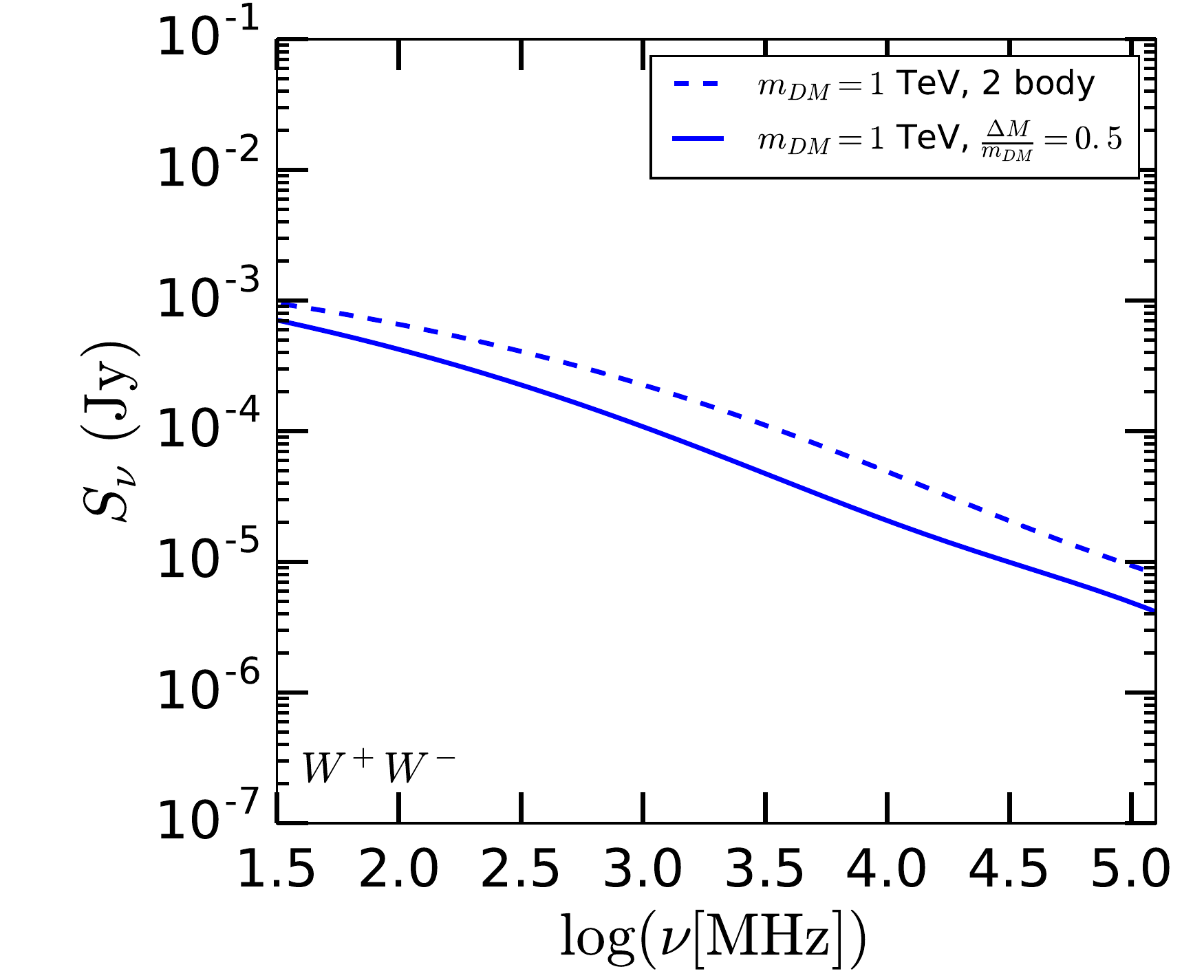}\hspace{2mm}%
  \includegraphics[height=0.4\textwidth, angle=0]{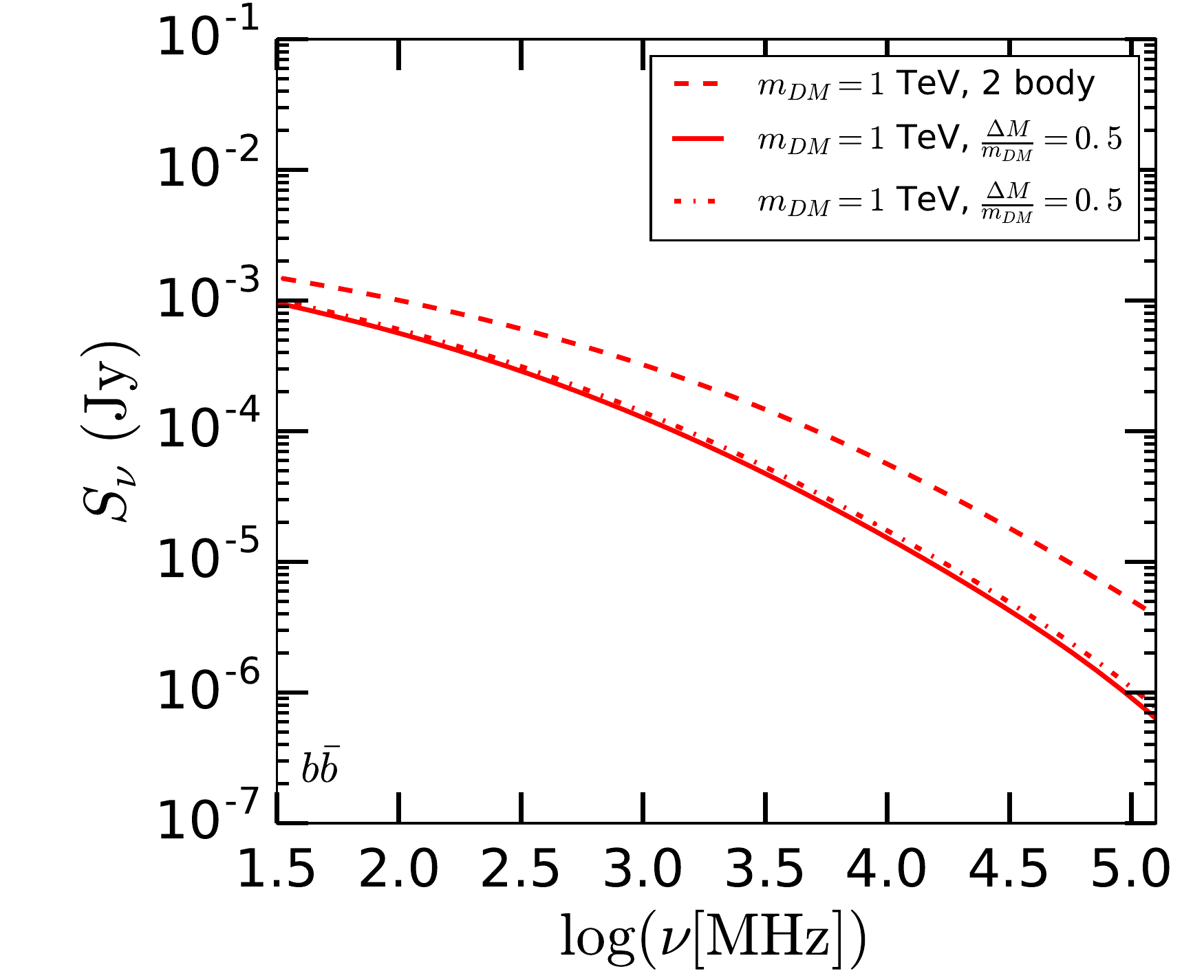}\hspace{2mm}%
  \caption{{\it Left panel:} Radio synchrotron flux, $S_{\nu}(\nu)$ (in Jy) for the decay processes $\phi \rightarrow W^+ W^-$ (dashed curve) 
  and $\phi_2 \rightarrow \phi_1 W^+ W^-$ (solid curve). In the later case, the (normalised) energy distribution of $W^+/W^-$ is governed by Eqn. \ref{eqn:diboson_dist}.
  {\it Right panel:} Synchrotron flux for the decay processes $\phi \rightarrow b\bar{b}$ (dashed curve) and $\phi_2 \rightarrow \phi_1 b\bar{b}$ (solid and dashed-dotted curves).
  For three-body decays the (normalised) energy distributions of $b/\bar{b}$ is governed by Eqns. \ref{eqn:ffbar_dist1} (solid line) and \ref{eqn:ffbar_dist2} (dashed-dotted line).
  Here $m_{DM} = 1$ TeV ($\frac{\Delta M}{m_{DM}}$ = 0.5 for multicomponent scenario) and $\Gamma = 10^{-28} s^{-1}$.
  Choices of the astrophysical parameters are $D_0 = 3 \times 10^{28} \mbox{cm}^2 \mbox{s}^{-1}$ and $B = 1$ $\mu G$.}
  \label{fig:WWSyncflux}
\end{figure*}

As an example we have shown the $e^{\pm}$ distribution ($\Gamma\frac{dN^{e}}{dE}$) produced in the 
decays $\phi\rightarrow\,W^+W^-$ ($\phi \rightarrow b\bar{b}$) and $\phi_2\rightarrow\phi_1W^+W^-$ ($\phi_2 \rightarrow \phi_1 b\bar{b}$) in the 
left (right) panel of Fig.~\ref{fig:WWelectronflux} for a benchmark value of $m_{DM}$,
$\Gamma$, $\Delta\,M/m_{DM}$. The energy distributions are softer for three-body decays, as in the case of $\gamma$-rays.
Fig.~\ref{fig:WWSyncflux} encapsulates the resulting synchrotron fluxes ($S_{\nu}(\nu)$) where the values of 
the diffusion coefficient($D_0$) and magnetic field($B$) have been chosen to be 
$D_0 = 3 \times 10^{28} \mbox{cm}^2 \mbox{s}^{-1}$ and $B = 1$ $\mu G$, for illustration \cite{Spekkens:2013ik,Colafrancesco:2006he}.

\section{Results}
\label{sec:limits}
Using Fermi-LAT observation of isotropic gamma-ray background (IGRB)~\cite{Ackermann:2014usa} we start by 
showing illustrative upper limits on the DM decay width $\Gamma$, 
considering only a single decay channel at a time (i) for a given dark matter mass ($m_{DM}$) in
case of two-body decays of DM itself and (ii) for a chosen dark matter mass ($m_{DM}$) and  two fixed values of $\Delta M/m_{DM}$, namely 0.9 (\textit{`hierarchical scenario'}) and 0.1 (\textit{`degenerate scenario'}) in case of 
three-body decays occurring within a multicomponent scalar dark sector. 
We have subsequently presented the upper limits on the Wilson coefficients 
in Eqns.~\ref{eqn:dim5_single} and \ref{eqn:dim6_multi} considering only one effective operator at a time. 
This is a reasonable assumption since each operator presented is 
independently gauge invariant. We have thus taken into account in the ultimate analysis all the decay 
channels opened up by a particular operator. The upper limits have been determined following the procedure of~\cite{Blanco:2018esa}.

In each of the above cases we have also presented the sensitivity reach of the upcoming Square Kilometre Array (SKA).
A large effective area and better baseline coverage help the SKA to achieve a significantly higher surface brightness 
sensitivity compared to existing radio telescopes\footnote{While estimating the predicted signal for SKA, we have assumed that 
the SKA field of view is larger than the dSph size considered here and hence 
all the flux from the dSph will contribute to the detected signal. This assumption may not be valid for the SKA precursors
like the Murchison Widefield Array (MWA) where the effect of the beam size needs to be accounted for while calculating the 
signal \cite{Kar:2019hnj}.}. We have used the documents provided in the SKA website \cite{SKA} for 
estimating the noise sensitivity. The higher sensitivity allows one to observe very low intensity radio signal coming from 
ultra-faint dSph's which are estimated to have sizable 
DM densities, and the radio synchrotron signals from them possibly have less 
astrophysical background, since they have low rates of star formation. For details of the analysis 
see reference \cite{Kar:2019cqo}.

To present our results for SKA we have assumed some benchmark values of the diffusion coefficient and the magnetic field, 
namely, $D_0 = 3 \times 10^{28} \mbox{cm}^2 \mbox{s}^{-1}$ and $B = 1$ $\mu G$. However, these astrophysical parameters are not very well constrained yet for a 
dSph \cite{Regis:2014koa,Natarajan:2013dsa}. 
Though the proximity to our galaxy suggests that $B \approx 1$ $\mu G$ is a reasonable possibility \cite{Regis:2014koa},
similar guidelines regarding $D_0$ hardly exists. Keeping this in mind, we have also shown the allowed astrophysical parameter space ($B-D_0$ plane) that 
can give rise to visible signal at SKA
when the particle physics parameters are set at benchmark values consistent with IGRB observation.   

\subsection{Limits on particle physics parameters}
\label{sec:limit_partphys}

\subsubsection{Decay to Gauge bosons}
\label{sec:limit_partphys_VV}

Fig.~\ref{fig:VV_G_mDM} shows the upper limits on the DM decay widths ($\Gamma$) from Fermi-LAT observations as well as the sensitivity reach of the SKA 
in the channels $W^{+}W^{-}$(upper panel, left), $ZZ$(upper panel, right), $Z\gamma$(lower panel, left) and $\gamma \gamma$ (lower panel, right). 
We have assumed 100\% branching ratio to each of these decay modes. It is important to point out that the decays of DM (or dark sector particles) 
to $Z\gamma$ and $\gamma \gamma$ are associated with primary (direct) photons. Since Fermi-LAT is mostly sensitive to photons in the energy 
range \textit{a few} MeV-1 TeV and the direct photons produced in the $Z\gamma$ and $\gamma \gamma$ final states for $m_{DM} > 1\,$TeV fall 
outside the energy range of Fermi-LAT, the corresponding limits weakens~\cite{Ackermann:2014usa}. 
The future generation gamma-ray experiment like CTA~\cite{Consortium:2010bc,Morselli:2017ree} can improve over Fermi-LAT in this range of 
parameters. We have adopted the strategy outlined in~\cite{Garny:2010eg} to calculate sensitivity reach of CTA in the channels 
$Z\gamma$ and $\gamma \gamma$, which will show up as \textit{sharp spectral features} on top of otherwise isotropic background 
flux of electrons+gamma-rays. The sensitivity reach of SKA in each of the cases have been shown which is nearly 4 to 2 orders of magnitude 
stronger depending on the DM mass $m_{DM}$  barring the $\gamma \gamma$ final state. In case of the $\gamma \gamma$ final state the 
primary (off-shell) photons can split into $e^+e^-$ pairs or other SM particle pairs which subsequently generate SKA-detectable radio signal~\footnote{
Unlike in the case of other SM final states, the $e^{\pm}$ spectrum that originates from the splitting of a virtual primary photon (in $\gamma\gamma$ and $\gamma Z$ 
final states) has been calculated using the tools provided in~\cite{Cirelli:2010xx,Ciafaloni:2010ti,Cirelli}.}. 
Since this splitting is suppressed by $\alpha_{\rm EM}$, the SKA sensitivity can be stronger by 2 to 1 order of magnitude only (see fig.~\ref{fig:VV_G_mDM} 
bottom panel, right). Better sensitivity of SKA is mostly attributed to its 
large cross-sectional area and low threshold \cite{SKA}. Of course, it also depends on the choice of astrophysical parameters 
($B, D_0$). In our case we have assumed that $D_0 = 3 \times 10^{28} \mbox{cm}^2 \mbox{s}^{-1}$ and $B = 1$ $\mu G$ which are 
reasonable choices for dSph such as Draco. A more conservative choice, i.e, a larger $D_0$ or a 
lower $B$ will raise the sensitivity level. 

The limits and sensitivities for the two-body decays $\phi \rightarrow VV^{\prime}$ is the strongest one since the final state $VV^{\prime}$ 
has the energy $m_{DM}$ available to them in these cases. For the three-body decays $\phi_2 \rightarrow \phi_1 VV^{\prime}$, on the other hand, 
even if one neglects the energy carried away by $\phi_1$ the energy available to $VV^{\prime}$ is $\approx \Delta M < m_{DM}$. Thus the energy 
distribution of the final state photons or $e^{\pm}$ softens for the three-body decays(see Fig.\ref{fig:WWgammaflux} and \ref{fig:WWelectronflux}). 
This explains why the limits weakens for three-body decays as compared to two-body decays and also by at least an order of magnitude in case of 
$\Delta M /m_{DM}=0.1$ compared to $\Delta M /m_{DM}=0.9$. The (normalised) energy distribution of $V/V'$, produced in the decay 
$\phi_2 \rightarrow \phi_1 VV^{\prime}$, is governed by the Eqn. \ref{eqn:diboson_dist}.

In Fig.~\ref{fig:fVV_mDM} we have shown the constraints obtained from Fermi-LAT and the sensitivity reach expected from SKA for gauge 
invariant wilson coefficients $f_{WW}$, $f_{BB}$ considering $\Lambda = 10^{16}$ GeV for illustration. The operator proportional to $f_{WW}$ opens up all 
the channels $W^{+}W^{-},ZZ,Z\gamma,\gamma \gamma$ while $f_{BB}$ opens up $ZZ,Z\gamma,\gamma \gamma$ (see Eqns. \ref{eqn:FVVp_expression1} 
and \ref{eqn:FVVp_expression2}). Thus in order to calculate limits on those parameters one needs to consider contribution from all the 
channels with appropriate branching fractions. As mentioned earlier, for channels such as $Z\gamma$ and $\gamma\gamma$ (which have direct 
photon(s) in their final states), both Fermi-LAT limit (mainly for lower DM mass) and CTA limit (mainly for 
higher DM mass) have been used. One should note that $\gamma \gamma$ channel has a branching ratio proportional to $\cos^4 \theta_W$ when $f_{BB}$ 
is open compared to the $\sin^4 \theta_W$ dependence, when $f_{WW}$ is open and thus the limit on $f_{BB}$ is affected more by the inclusion of 
the sensitivity reach of CTA. This understanding is reflected in the kink around the energy, beyond which CTA offers a better probe for direct 
photons than Fermi-LAT. We have also shown the limits on the wilson coefficient $f_B$. The operator proportional to $f_B$ gives only a two-body 
decay $\phi_2 \rightarrow \phi_1 Z$ (see the discussions regarding Eqn. \ref{eqn:p2p1V_eff_opp}).

\subsubsection{Decay to Fermions}
\label{sec:limit_partphys_ff}

In Fig.~\ref{fig:FF_G_mDM} we have shown the upper limits (for Fermi-LAT) and sensitivity (for SKA) on the decay width ($\Gamma$) as a function 
of $m_{DM}$ assuming the decay occurs dominantly through $b\bar{b}$ (upper panel, left), $t\bar{t}$ (upper panel, right) and $\tau^+\tau^-$ (lower panel). 
The (normalised) energy distribution of a fermion($f$)/anti-fermion($\bar{f}$) produced in the three-body decay $\phi_2 \rightarrow \phi_1 f\bar{f}$ is considered 
to be the one governed by Eqn.~\ref{eqn:ffbar_dist1}. One can check that the other energy distribution provided in Eqn.~\ref{eqn:ffbar_dist2} produces almost similar 
limits (and sensitivity) for any of the aforementioned fermionic channels, as expected from Figs.~\ref{fig:WWgammaflux} and \ref{fig:WWelectronflux}. 

In Fig.~\ref{fig:fFF_mDM} we have shown the constrains obtained from Fermi-LAT and the sensitivity reach expected from SKA on the wilson coefficient 
$f_{qq}$ in case of the two-body decay of DM itself (upper left panel) and in case of decays in the dark sector (lower left panel). Here $f_{qq}$ 
is the coupling to the quarks arising from $\mathcal{L}^{\rm fermion,1}_{dim-5}$ in Eqn. \ref{eqn:dim5_single} and 
$\mathcal{L}^{\rm fermion,1}_{dim-6}$ \ref{eqn:dim6_multi}. For simplicity we have assumed that both the left and right handed quarks have 
the same values of this coupling. Also, we have considered that $f_{qq} \ne 0$ only for third generation of quarks 
which via gauge invariance dictates the relative contribution of the channels $b\bar{b}$ and $t\bar{t}$. 
It is quite evident that the limits from SKA on $f_{qq}$ will be stronger than Fermi-LAT by more than one order of 
magnitude up to $m_{DM} \simeq 100\,$TeV~\footnote{It may be noted that DM decay takes place via higher dimensional operators 
with a large suppression scale. Thus one does not expect any unitarity bounds on the mass of decaying DM. On the other hand, 
such bounds may restrict $m_{\rm DM}$ to be less than few tens of TeV from the viewpoint of annihilations \cite{PhysRevLett.64.615}, 
on which we have not entered into a discussion here.}.

The upper right and lower right panel of the same figure shows the constraints on the wilson coefficient $f_{ll}$ in case of the two-body 
decay of DM itself and in case of decays in the dark sector, respectively. Here also, we have assumed that both the left and right handed 
leptons have the same values of this coupling which appears only for the third generation of leptons. The only visible decay products as a 
result of switching on $f_{ll}$ being $\tau^{+}\tau^{-}$ the limits on $f_{ll}$ is 
straightforward to obtain from the decay widths themselves. Although the sensitivity of 
SKA to $\tau^{+}\tau^{-}$ final states decreases rapidly as the $m_{DM}$ increases (see the lower panel of 
Fig.\ref{fig:FF_G_mDM})\footnote{In case of $\tau^{+}\tau^{-}$ channel one mostly has high energy $e^{\pm}$ which give
rise to a synchrotron flux peaking towards higher frequencies. Thus for heavier DM masses only the lower frequency part of 
the radio flux (which is suppressed) contributes to SKA observation and consequently sensitivity decreases with 
increasing $m_{\rm DM}$ (See \cite{Kar:2019cqo} for more details). }, SKA can still probe larger parameter space 
compared to Fermi-LAT even up to $m_{DM} \simeq 100\,\,$TeV. 

Fig.~\ref{fig:fFFH_mDM} shows the constrains obtained from Fermi-LAT and the sensitivity reach expected from SKA on the 
wilson coefficients $f_{bbH}$, $f_{ttH}$ and $f_{llH}$ in case of the two-body decay of DM itself (left column) and in case of decays in the 
dark sector (right column). Here $f_{bbH}$, $f_{ttH}$ and $f_{llH}$ are the couplings to the $b$ and $t$ quarks and $\tau$ lepton following 
from $\mathcal{L}^{\rm fermion,2}_{dim-5}$ in Eqn. \ref{eqn:dim5_single} and $\mathcal{L}^{\rm fermion,2}_{dim-6}$ in Eqn.~\ref{eqn:dim6_multi}. For simplicity, 
we have considered the couplings to be non-zero only for the third generation of fermions. It is clear that the limits expected from 500 hours of observation at SKA 
on $f_{bbH}$ and $f_{ttH}$ will be stronger than Fermi-LAT by more than one order of magnitude up to $m_{DM} \simeq 100\,$TeV. Although the sensitivity of 
SKA to the $\tau^{+}\tau^{-}$ channel decreases rapidly as $m_{DM}$ increases (see the lower panel of Fig. \ref{fig:FF_G_mDM}), SKA can still probe larger parameter 
space compared to Fermi-LAT even up to $m_{DM} \simeq 100\,\,$TeV.

\subsection{Limits on Astrophysical parameters}
\label{sec:limit_astrophys}

As of now we have shown the particle physics parameter space that can be probed using SKA choosing a benchmark values of 
$D_0 = 3 \times 10^{28} \mbox{cm}^2 \mbox{s}^{-1}$ and $B = 1$ $\mu G$. Here we have shown which region of the astrophysical parameters
$B-D_0$ will produce a observable DM decay signal at SKA assuming a set of particle physics parameters that has not been ruled out by
isotropic $\gamma$-ray background (IGRB) observation. 

In Fig.~\ref{fig:D0_B_fVV} we have considered either only $f_{WW}$ is non-zero (left panel) or only $f_{BB}$ is 
non-zero (right panel) and chosen a benchmark value $m_{DM}= 10\,$TeV while the values of the wilson coefficients are dictated by the 
corresponding upper limits as obtained from Fermi-LAT (and in case of $Z\gamma$ and $\gamma\gamma$ channels by CTA) as shown in Fig. \ref{fig:fVV_mDM}. The 
regions above the curves shown in the figures are favourable for observation in SKA (assuming a 500 hours of observation). The corresponding limits on 
$B - D_0$ for $f_B \neq 0$ can be obtained similarly.
The results for $f_{qq}$ (Fig.~\ref{fig:D0_B_fffbarderiv}, left panel) , $f_{ll}$ (Fig.~\ref{fig:D0_B_fffbarderiv}, right panel) 
as well as $f_{bbH}$ (Fig.~\ref{fig:D0_B_fffbarHderiv}, left panel) and  $f_{llH}$ (Fig.~\ref{fig:D0_B_fffbarHderiv}, right panel) should also be interpreted 
in the same way. Note that if one considers only the anti-galactic contribution to the galactic $\gamma$-ray flux in deriving the $\gamma$-ray constraints
on the Wilson coefficients (as discussed in sec. \ref{sec:gammasignals}), the limits on the $B-D_0$ plane presented here can become stronger at most by a factor 
of 5. 

Our general conclusion emerges from Figs.~\ref{fig:D0_B_fVV}-\ref{fig:D0_B_fffbarHderiv}. We demonstrate that the SKA can do considerably better than $\gamma$-ray 
observations for the range of $m_{\rm DM}$ under focus here. If SKA indeed records such radio signals as predicted here, then, some independent information on 
$m_{\rm DM}$ and $\Gamma$ may enable one to identify regions in the $B-D_0$ plane, which are consistent with the DM decay observations. Similar conclusions 
involving DM annihilations in a dSph can be found in \cite{Kar:2019cqo}.   

\section{Conclusion}
\label{sec:Conclusion}
We have carried out a study of long-lived DM with its decay showing
up in $\gamma$-ray as well as radio telescope observations. In order to
comply with constraints on DM lifetime, the decay interactions have been
parametrised by higher-dimensional operators. Both two-body decays
of a scalar DM particle and three-body decays of quasi-stable particles
within a dark sector have been considered, the SM particles among decay products
being pairs of gauge bosons as well as fermions of the third family.

Constraints on the coefficients of the various operators have been obtained from
existing $\gamma$-ray observations. The Fermi-LAT results are found to
be most constraining in this respect, In comparison, the proposed CTA observations are found to yield weaker constraints, 
except in cases where at least one $\gamma$-ray photon is directly produced in decay, as opposed to photons coming via cascades. 

However, radio synchrotron signals from dSph's are found to provide better probes into DM decays, by enabling exploration of 
regions which cannot be ruled out by
either the Fermi-LAT data or the CTA. This is true even for DM masses
well-above a TeV. Using as benchmark 500 hours of observation at the upcoming
SKA radio telescope,  we find such a conclusion to hold for DM masses 
ranging up to tens of TeV.

It is also shown how some independent conclusion on DM  mass
and its decay rate can enable one to identify viable regions of astrophysical
parameters pertaining to a dSph under observation. In this spirit,
we demonstrate how to find allowed ranges in the space spanned by the 
diffusion coefficient in the dSph and the galactic magnetic field,
for sample values of the DM particle mass and its decay rate,
consistent with $\gamma$-ray observation.

\section{Acknowledgements}
The authors thank Alejandro Ibarra, Saurav Mitra, Tirthankar  Roy Choudhury and Steven Tingay for their help in learning about
gamma-ray and radio signals of dark matter.
AG and AK acknowledges the hospitality of the Theoretical
Physics Department, Indian Association for Cultivation of Science(IACS), Kolkata and Indian 
Institute of Science Education and Research(IISER), Kolkata where a substantial part of the project was carried out. 
This work was partially supported by funding available from the Department of Atomic Energy, Government of India, for the 
Regional Centre for Accelerator-based Particle Physics (RECAPP), 
Harish-Chandra Research Institute, Allahabad.
\appendix 
\section{Single-component dark matter}
\label{AppendixA}
\subsection{Decay to gauge bosons}
The decay $\phi\rightarrow\,V\,V^{\prime}$ is governed by the 
term 
\begin{equation}
\dfrac{f_{VV^{\prime}}}{\Lambda}\phi\,V_{\mu\nu}V^{\prime\mu\nu}
\end{equation}
and the partial width is,
\begin{eqnarray}
\Gamma_{VV^{\prime}}&=&\frac{1}{2\pi}\left(\frac{f_{VV^{\prime}}}{\Lambda}\right)^2\,M^3\sqrt{1-\frac{2(m^2_V+m^2_{V^\prime})}{M^2}+\frac{(m^2_V-m^2_{V^{\prime}})^2}{M^4}}\nonumber\\
&&\left[1-\frac{2(m^2_V+m^2_{V^\prime})}{M^2}+\frac{(m^4_V+m^4_{V^{\prime}}+4m^2_Vm^2_{V^{\prime}})}{M^4}\right].
\label{eqn:VV_width}
\end{eqnarray}

Here,
\begin{align}
f_{VV^{\prime}}= 
\begin{cases}
\displaystyle{f_{WW}}&~~{\rm for}\, V=W^{+},V^{\prime}=W^{-}
\\
\displaystyle{\frac{1}{\sqrt{2}}\left(\cos^2\theta_{W}f_{WW}+\sin^{2}\theta_{W}f_{BB}\right)}&~~{\rm for}\, V=Z,V^{\prime}=Z
\\
\displaystyle{\cos\theta_{W}\sin\theta_{W}\left(f_{WW}-f_{BB}\right)}&~~{\rm for}\, V=Z,V^{\prime}=\gamma
\\
\displaystyle{\frac{1}{\sqrt{2}}\left(\sin^2\theta_{W}f_{WW}+\cos^{2}\theta_{W}f_{BB}\right)}&~~{\rm for}\, V=\gamma,V^{\prime}=\gamma
\end{cases}\;.
\label{eqn:FVVp_expression1}
\end{align}

\subsection{Decay to fermions}
The decay of $\phi$ to fermion pairs is parametrized as,
\begin{equation}
\frac{f_{ff}}{\Lambda}\phi\,\left(\bar{f}_{L}\gamma^{\mu}{\partial}_{\mu}f_{L}+\bar{f}_{R}\gamma^{\mu}{\partial}_{\mu}f_{R}\right)+\frac{f_{ffH}}{\Lambda}\phi\left[\bar{f}_{L}f_{R}H+h.c\right]
\label{eqn:ff_eff_opp_single}
\end{equation}
The decay width from the first term in Eqn.~\ref{eqn:ff_eff_opp_single},
\begin{eqnarray}
\Gamma_{f\bar{f}}&=&\,N_c\,\frac{1}{8\pi}\left(\frac{f_{ff}}{\Lambda}\right)^2\,m^2_f\,M\,\left(1-\frac{4m^2_f}{M^2}\right)^{3/2}.
\end{eqnarray} 
and from the second term in Eqn.~\ref{eqn:ff_eff_opp_single},
\begin{eqnarray}
\Gamma_{f\bar{f}}&=&\,N_c\,\frac{1}{16\pi}\left(\frac{f_{ffH}}{\Lambda}\right)^2\,v^2\,M\,\left(1-\frac{4m^2_f}{M^2}\right)^{3/2}.
\end{eqnarray} 
where $N_c=3$ for quarks and $N_c=1$ for leptons.

\section{Multicomponent dark sector}
\label{AppendixB}
\subsection{Decay to gauge bosons}
\begin{enumerate}
\item
The decay $\phi_{2}\rightarrow\,\phi_{1}\,V\,V^{\prime}$ is governed by
\begin{equation}
\dfrac{f_{VV^{\prime}}}{\Lambda^{2}}\phi_{2}\phi_{1}V_{\mu\nu}V^{\prime\mu\nu}
\label{eqn:VV_eff_opp}
\end{equation}

The energy distribution of the vector boson $V$ originating from 
$\phi_2\rightarrow\phi_1\!VV^{\prime}$ is,
\begin{eqnarray}
\frac{d\Gamma_{\phi_1\,VV^{\prime}}}{dx_v}\,&=&\,\left(\frac{f_{VV^{\prime}}}{\Lambda^{2}}\right)^{2}\,\frac{M^{5}_{2}}{32\pi^{3}}\sqrt{x^{2}_{v}-4r^{2}_{v}}\,\lambda^{1/2}\left(\sqrt{1+r^{2}_{v}-x_v}\,,\,r_{1},\,r_{v^{\prime}}\right)\nonumber\\
&&\bigg[2r^{2}_{v}r^{2}_{v^{\prime}}+\frac{1}{12}\bigg\{3\frac{(x_{v}-2r^2_v)^{2}}{(1+r^{2}_{v}-x_v)^{2}}(1-r^{2}_{1}+r^{2}_{v}+r^{2}_{v^{\prime}}-x_{v})^{2}\nonumber\\
&&\,+(x^{2}_v-4r^{2}_{v})\lambda\left(\sqrt{1+r^{2}_{v}-x_v}\,,\,r_{1},\,r_{v^{\prime}}\right)\bigg\}\bigg]\nonumber\\
\label{eqn:diboson_dist}
\end{eqnarray}
where $r_{v(v^\prime)}=\dfrac{m_{v(v^\prime)}}{M_{2}}\,,\,r_{1}=\dfrac{M_{1}}{M_{2}}\,,\,x_{v}=
\dfrac{2E_{v}}{M_{2}}$ and $2\,r_{v}\,\le\,x_v\,\le\,\left(1-r^{2}
_{1}+r^2_v-r^2_{v^{\prime}}-2r_{1}r_{v^\prime}\right)$.\\
The kallen-lambda function is given by, $\lambda\left(a,b,c\right)\,=\,
\left(1-\dfrac{2\left(b^{2}+c^{2}\right)}{a^{2}}+\dfrac{\left(b^{2}-c^{2}\right)^{2}}{a^{4}}\right)$.\\

Here, 
\begin{align}
f_{VV^{\prime}}= 
\begin{cases}
\displaystyle{f_{WW}}&~~{\rm for}\, V=W^{+},V^{\prime}=W^{-}
\\
\displaystyle{\frac{1}{\sqrt{2}}\left(\cos^2\theta_{W}f_{WW}+\sin^{2}\theta_{W}f_{BB}\right)}&~~{\rm for}\, V=Z,V^{\prime}=Z
\\
\displaystyle{\cos\theta_{W}\sin\theta_{W}\left(f_{WW}-f_{BB}\right)}&~~{\rm for}\, V=Z,V^{\prime}=\gamma
\\
\displaystyle{\frac{1}{\sqrt{2}}\left(\sin^2\theta_{W}f_{WW}+\cos^{2}\theta_{W}f_{BB}\right)}&~~{\rm for}\, V=\gamma,V^{\prime}=\gamma
\end{cases}\;.
\label{eqn:FVVp_expression2}
\end{align}
\item Due to angular momentum conservation the decay $\phi_2 \rightarrow \phi_1 \gamma$ is forbidden
and the operator  
\begin{equation}
\dfrac{f_{B}}{\Lambda^{2}}\left(\partial_\mu \phi_{2} \partial_\nu \phi_{1} -\partial_\nu \phi_{2} \partial_\mu \phi_{1} \right)B_{\mu\nu}
\label{eqn:p2p1V_eff_opp}
\end{equation}
can only trigger the decay $\phi_2\rightarrow\phi_1 Z$ .  

The emitted $Z$-boson has a fixed energy $E_z=M_2/2 \left(1-r^2_1+r^2_z\right)$ and the corresponding width is,
\begin{equation}
\Gamma_{\phi_1 Z}= \left(\frac{f_B}{\Lambda^2}\right)^2 M^3_2 \frac{\sin^2\theta_W  m^2_z}{16 \pi} \lambda^{3/2}\left(1,r_1,r_z\right)
\end{equation}

\end{enumerate}

\subsection{Decay to fermions}
The interactions are,
\begin{eqnarray}
\frac{f_{ff}}{\Lambda^{2}}\phi_{2}\overset{\leftrightarrow}{\partial}_{\mu}\phi_{1}\,\bar{f}\gamma^{\mu}\,f+\frac{f_{ffH}}{\Lambda^{2}}\frac{v}{\sqrt{2}}\phi_{2}\phi_{1}\,\bar{f}\,f
\label{eqn:ff_eff_opp}
\end{eqnarray}
For simplicity in our analysis we have taken only one operator at a 
time. 

\begin{enumerate}
\item 
The differential distribution of the fermion due to first term of 
Eqn.~\ref{eqn:ff_eff_opp}.
\begin{eqnarray}
\frac{d\Gamma_{\phi_1\!f\bar{f}}}{dx}\,&=&\,N_c\,\left(\frac{f_{ff}}{\Lambda^{2}}\right)^{2}\,\frac{M^{5}_{2}}{16
\,\pi^{3}}\bigg[\sqrt{x^{2}-4\,r^{2}_{f}}\,\lambda^{1/2}\left(\sqrt{1+r^2_f-x},r_1,r_f\right)\nonumber\\
&&\frac{2(1-r^2_1-x)(1+r^2_f-x)-(1-x)(2-x)(1-r^2_1+2r^2_f-x)}{\left(1+r^{2}_{f}-x\right)}\bigg]
\label{eqn:ffbar_dist1}
\end{eqnarray}
where $r_f=\dfrac{m_f}{M_2}$,$r_1=\dfrac{M_1}{M_2}$ and $x=\dfrac{2E_f}{M_2}$ with $2r_f\leq\!x\leq\!\left(1-r^2_1-2r_1r_f\right)$.\\  

\item
The differential energy distribution of the fermions for the second 
operator in Eqn.~\ref{eqn:ff_eff_opp} is,
\begin{eqnarray}
\frac{d\Gamma_{\phi_1\!f\bar{f}}}{dx}\,&=&\,N_c\,\left(\frac{f_{ffH}}{\Lambda^{2}}\right)^{2}\,\frac{v^{2}M^{3}_{2}}{512\pi^{3}}\bigg[\sqrt{x^{2}-4\,r^{2}_{f}}\,\lambda^{1/2}\left(\sqrt{1+r^2_f-x},r_1,r_f\right)\nonumber\\
&&\frac{x(1-r^2_1-x)+2r^2_f(-3+r^2_1+4x)-8r^4_f}{\left(1+r^{2}_{f}-x\right)}\bigg]
\label{eqn:ffbar_dist2}
\end{eqnarray}
where $r_{f}=\dfrac{m_{f}}{M_{2}}\,,\,r_{1}=\dfrac{M_{1}}{M_{2}}\,,\,x=\dfrac{2E_{f}}{M_{2}}$ and $2\,r_{f}\,\le\,x\,\le\,\left(1-r^{2}
_{1}-2r_{1}r_{f}\right)$.\\

Here $N_c=3$ for quarks and $N_c=1$ for leptons.

\end{enumerate}


\providecommand{\href}[2]{#2}
\addcontentsline{toc}{section}{References}
\bibliographystyle{JHEP}

\bibliography{biblio}

\newpage

\begin{figure*}[ht!]
\centering
  \includegraphics[height=0.4\textwidth, angle=0]{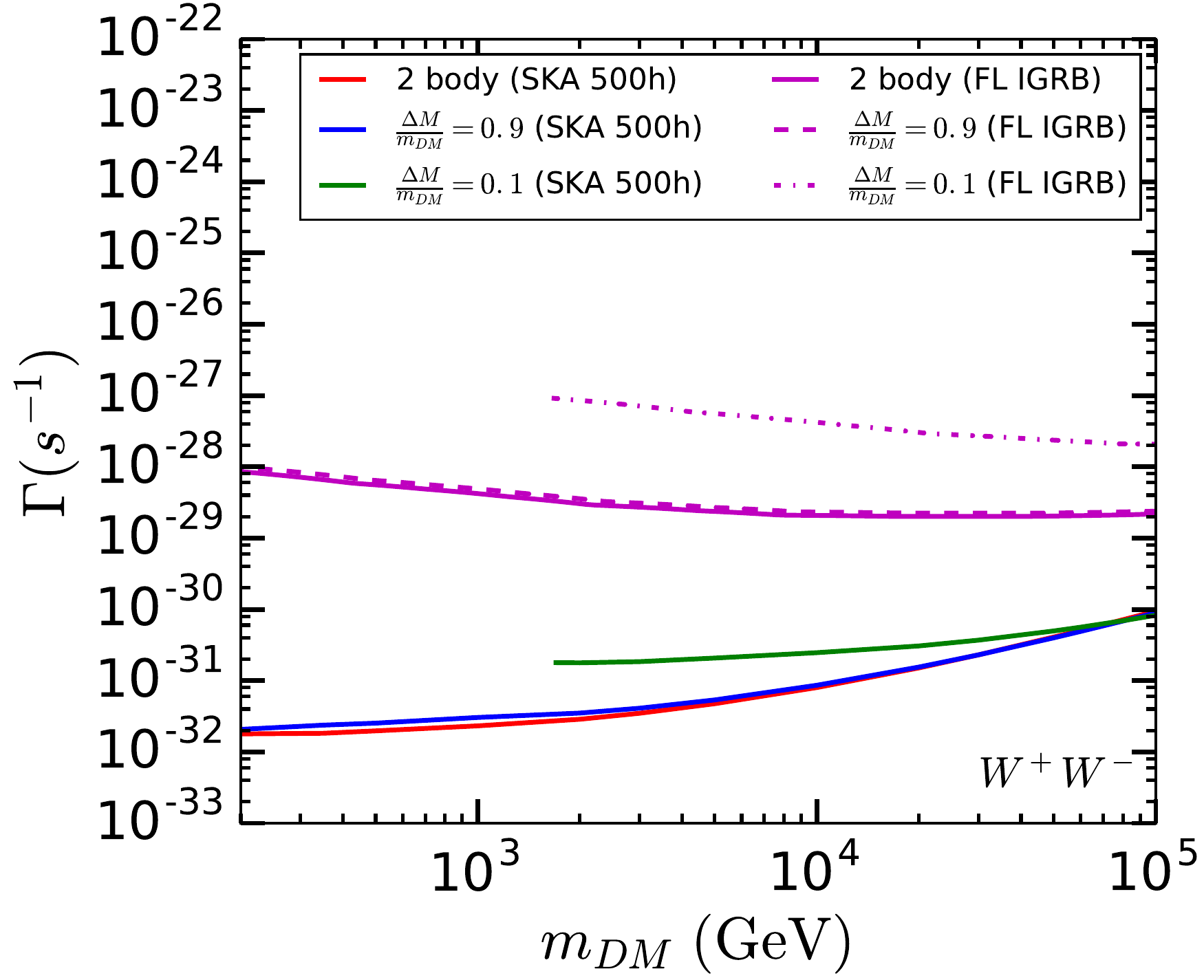}\hspace{2mm}%
  \includegraphics[height=0.4\textwidth, angle=0]{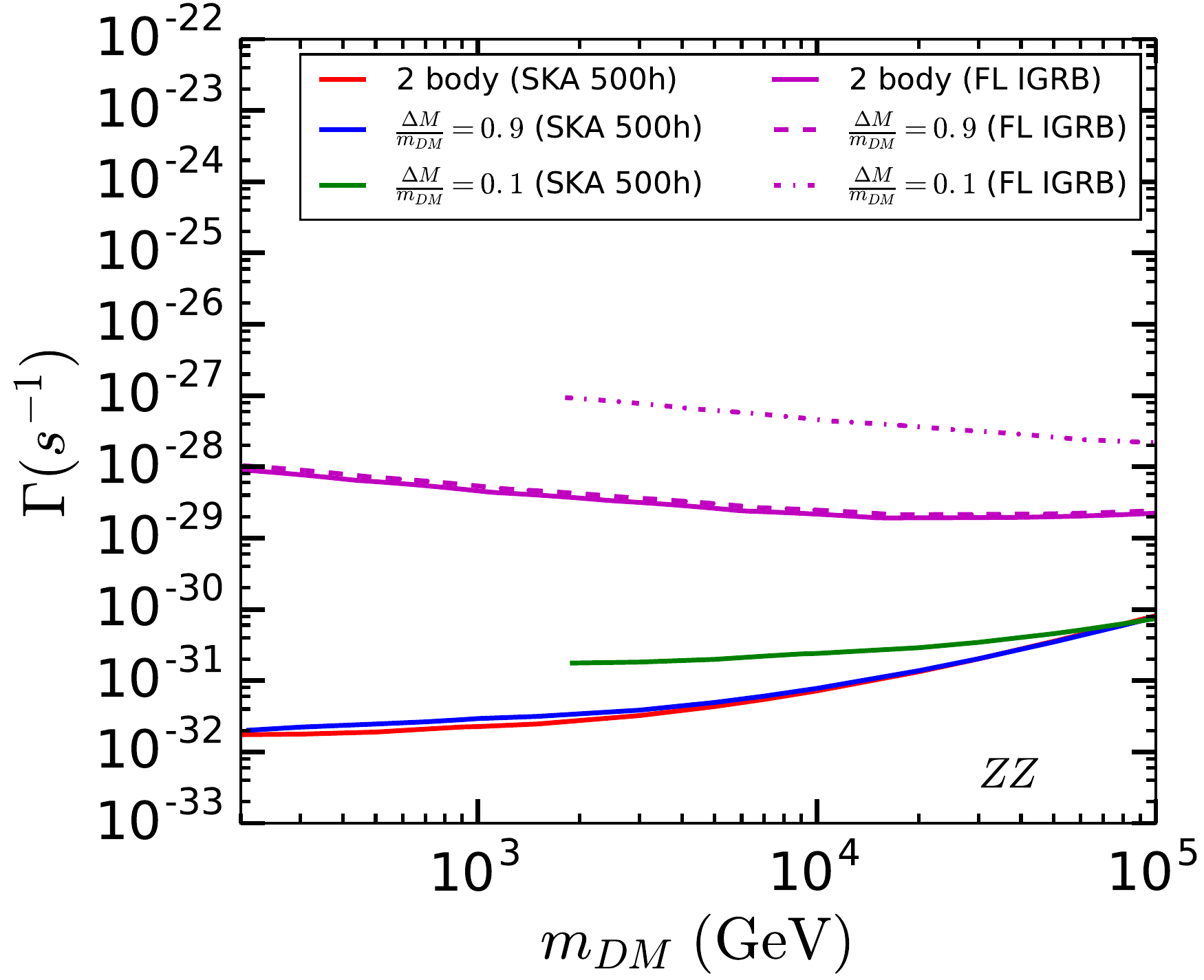}\hspace{2mm}%
  \includegraphics[height=0.4\textwidth, angle=0]{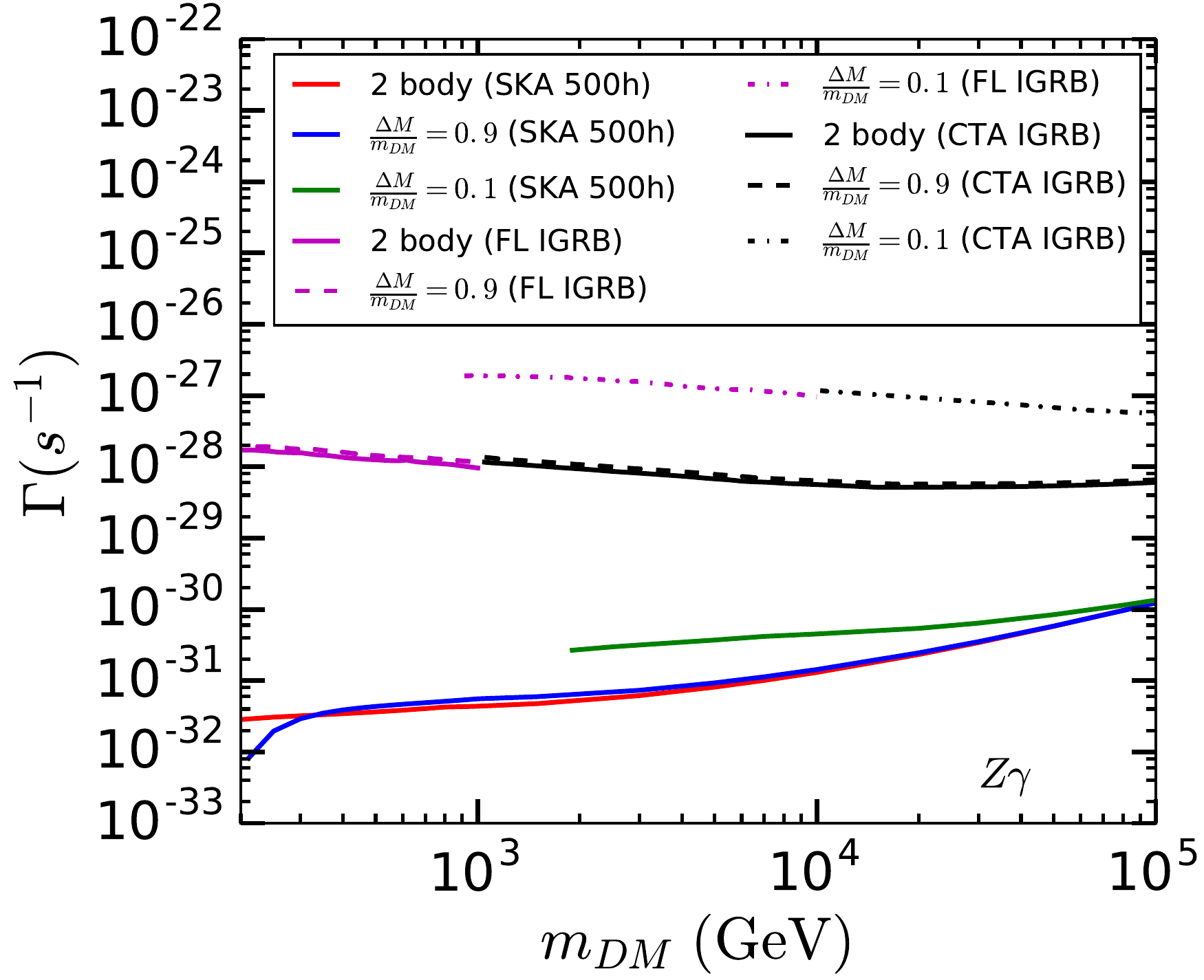}\hspace{2mm}%
  \includegraphics[height=0.4\textwidth, angle=0]{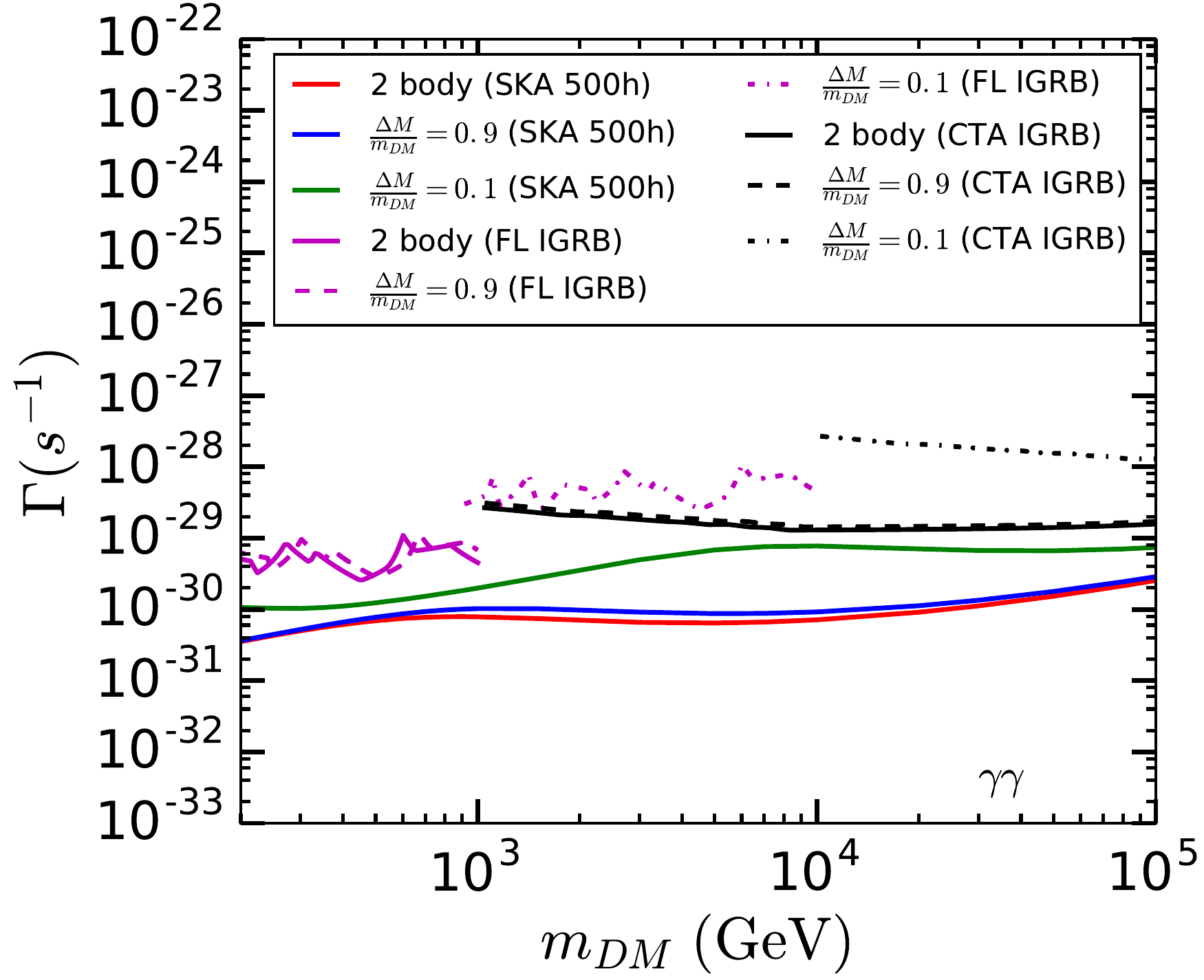}\hspace{2mm}%

  \caption{SKA sensitivity and the upper limit from $\gamma$-ray observation in the $\Gamma - m_{DM}$ plane for the decay of a DM (or a dark sector) particle of mass 
  $m_{DM}$ to various vector boson final states ($VV'$), i.e. either $W^+W^-$ (upper-left panel) or $ZZ$ (upper-right panel) or $Z\gamma$ 
  (lower-left panel) or $\gamma\gamma$ (lower-right panel). The red, blue and green curves denote the required values of $\Gamma$ to detect the 
  radio fluxes at SKA (assuming a 500 hours of observation) from Draco dSph for the processes $\phi \rightarrow VV'$ and $\phi_2 \rightarrow \phi_1 VV'$ 
  (with $\frac{\Delta M}{m_{DM}} = 0.9$ and 0.1), respectively. The (normalised) energy distribution of $V/V'$ in three-body decay is governed by 
  Eqn.~\ref{eqn:diboson_dist}. The astrophysical parameters used are $D_0 = 3 \times 10^{28} \mbox{cm}^2 \mbox{s}^{-1}$ and $B = 1$ $\mu G$. 
  The solid, dashed and dashed-dotted magenta lines are the corresponding upper limits on $\Gamma$ from the observation of isotropic 
  $\gamma$-ray background (IGRB) by Fermi-LAT (FL). In case of $Z\gamma$ and $\gamma\gamma$, along with Fermi-LAT, 
  the projected sensitivity from the IGRB observation by CTA (black curves; assuming 500 hours of observation) also has been shown (see the text for details).}
\label{fig:VV_G_mDM}
\end{figure*}

\begin{figure*}[ht!]
\centering
  \includegraphics[height=0.4\textwidth, angle=0]{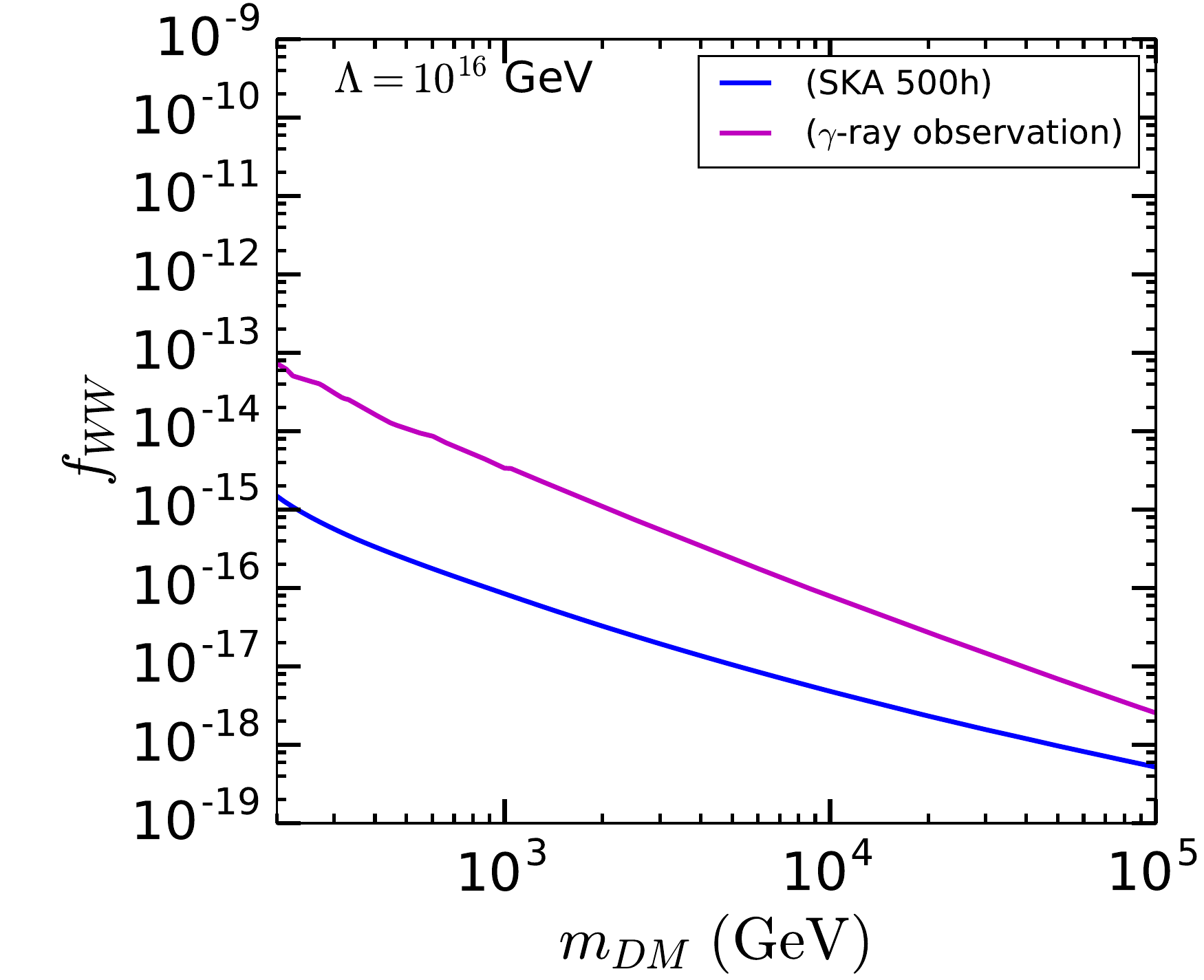}\hspace{2mm}%
  \includegraphics[height=0.4\textwidth, angle=0]{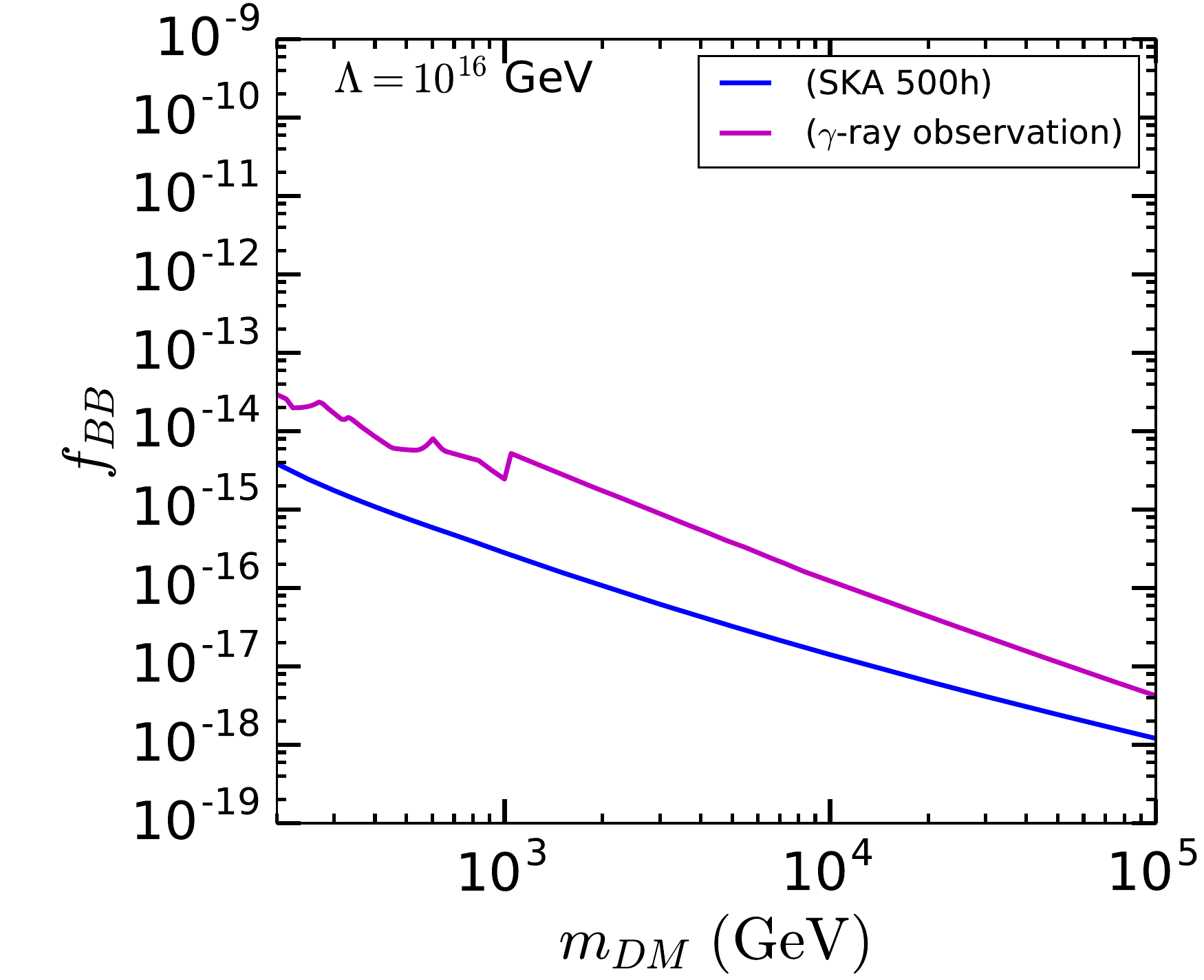}\hspace{2mm}%
  \includegraphics[height=0.4\textwidth, angle=0]{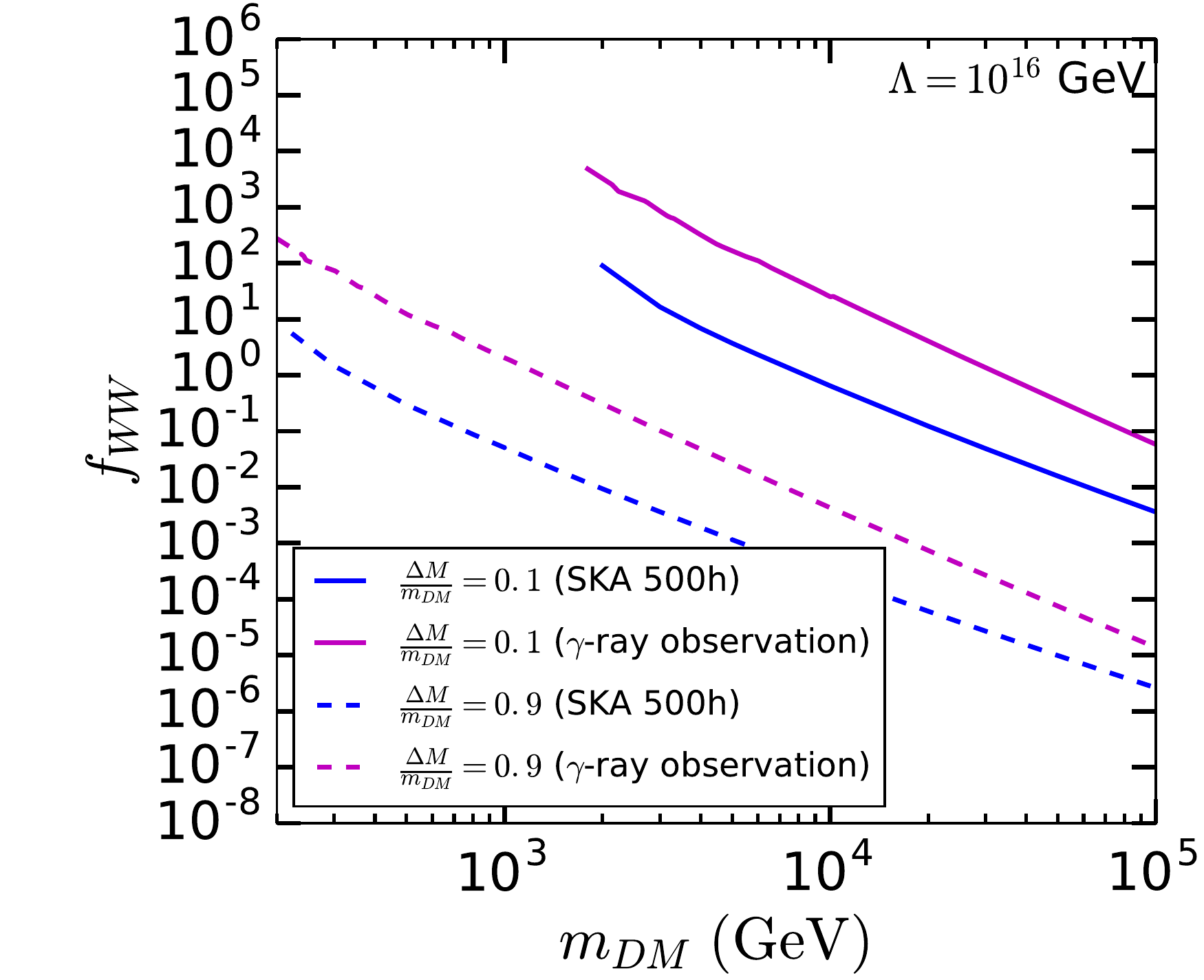}\hspace{2mm}%
  \includegraphics[height=0.4\textwidth, angle=0]{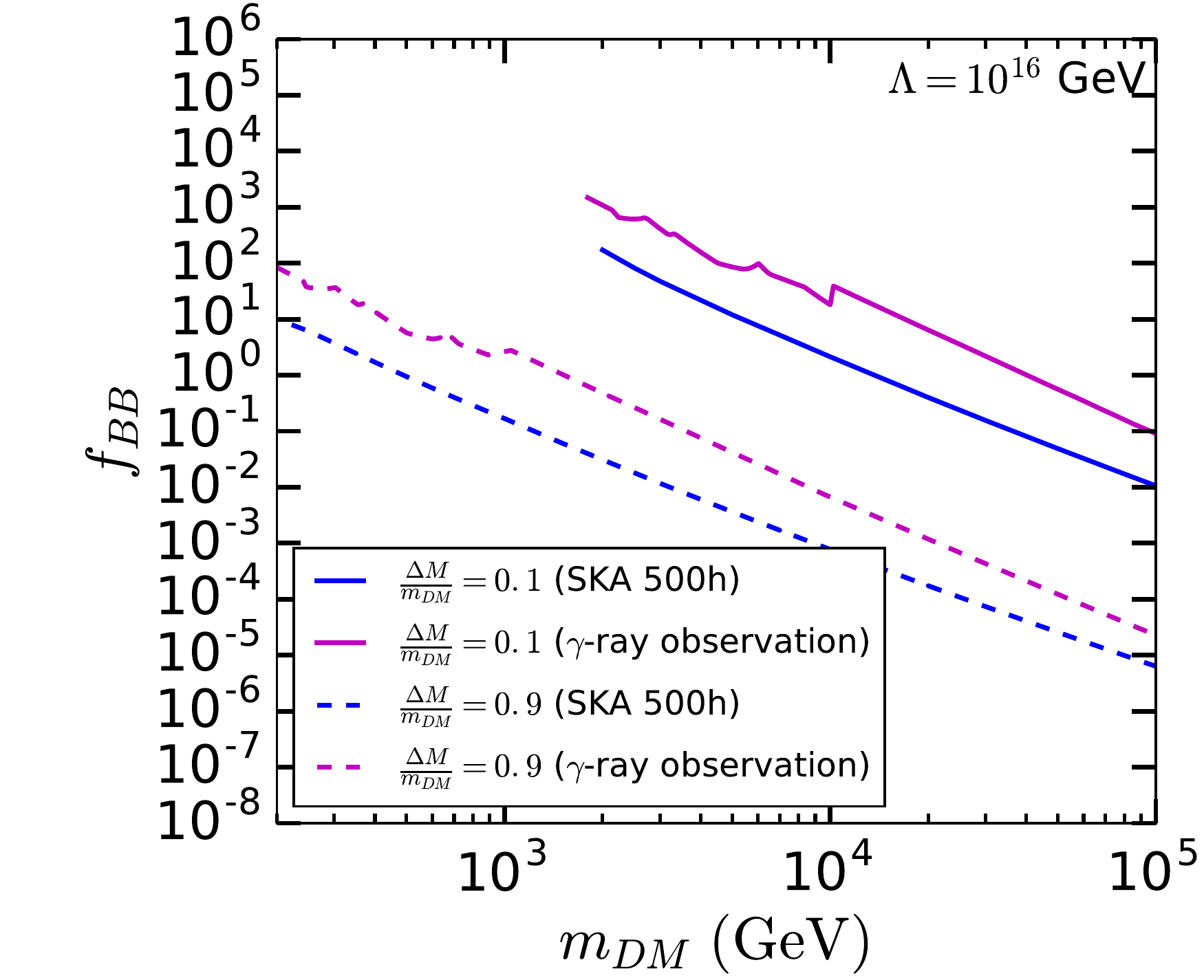}\hspace{2mm}%
  \includegraphics[height=0.4\textwidth, angle=0]{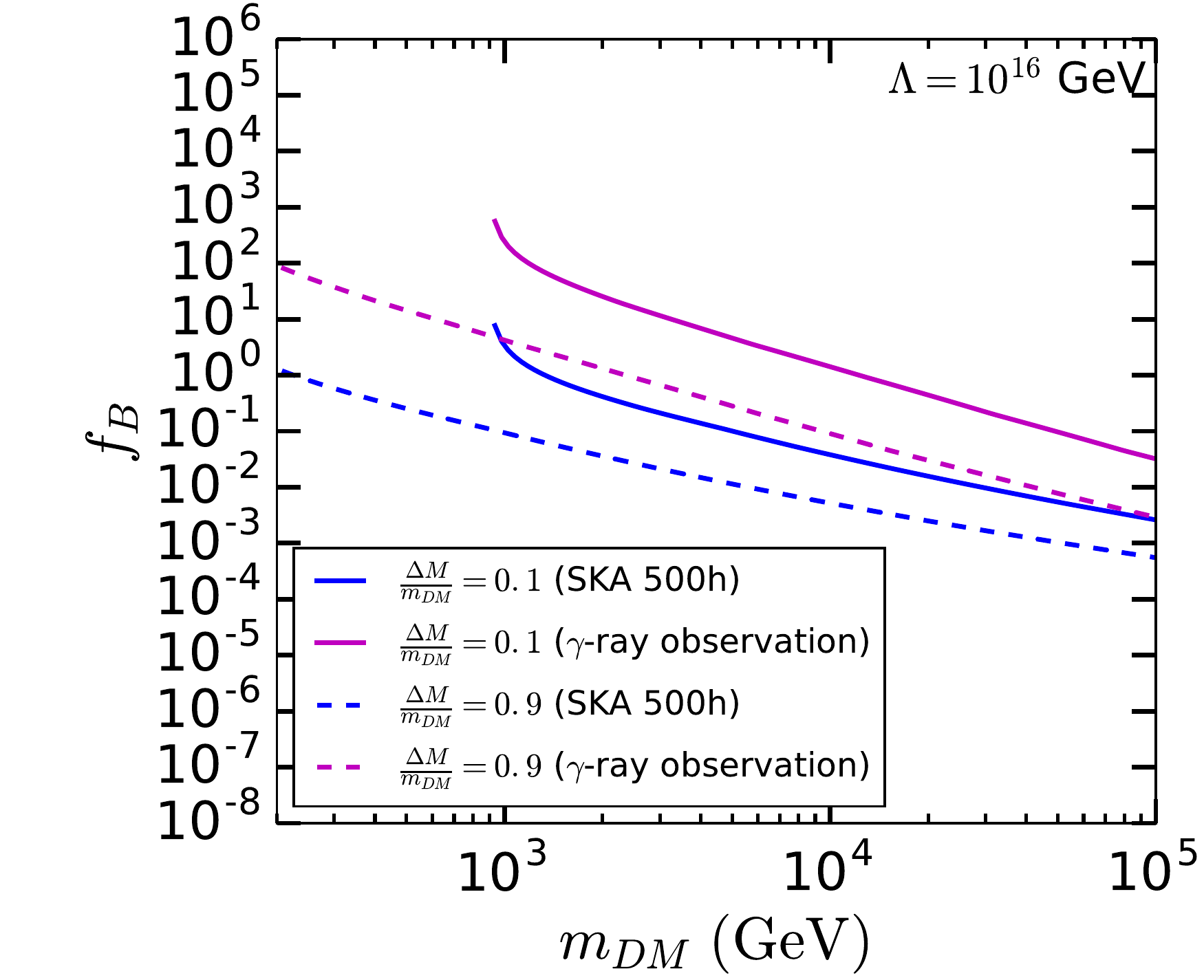}\hspace{2mm}%

  \caption{{\it Upper panel:} Sensitivity of SKA (blue lines; assuming a 500 hours of observation for Draco dSph) to various gauge boson 
  effective couplings (listed in Eqn. \ref{eqn:dim5_single}) which governs the two-body decays of a DM particle of mass $m_{DM}$ to the pairs 
  of gauge boson final states. The astrophysical parameters used are $D_0 = 3 \times 10^{28} \mbox{cm}^2 \mbox{s}^{-1}$ and $B = 1$ $\mu G$. 
  The magenta lines are the corresponding limits obtained from 
  the isotropic $\gamma$-ray background (IGRB) observation. {\it Middle and lower panels:} Similar constraints on the gauge boson 
  effective couplings (listed in Eqn. \ref{eqn:dim6_multi}) which give rise to the three-body decays of a DM particle of mass $m_{DM}$. Two different 
  values of $\Delta M /m_{DM}$ have been considered, $\Delta M /m_{DM} = 0.1$ (solid lines) and 0.9 (dashed lines).}
\label{fig:fVV_mDM}
\end{figure*}

\begin{figure*}[ht!]
\centering
  \includegraphics[height=0.4\textwidth, angle=0]{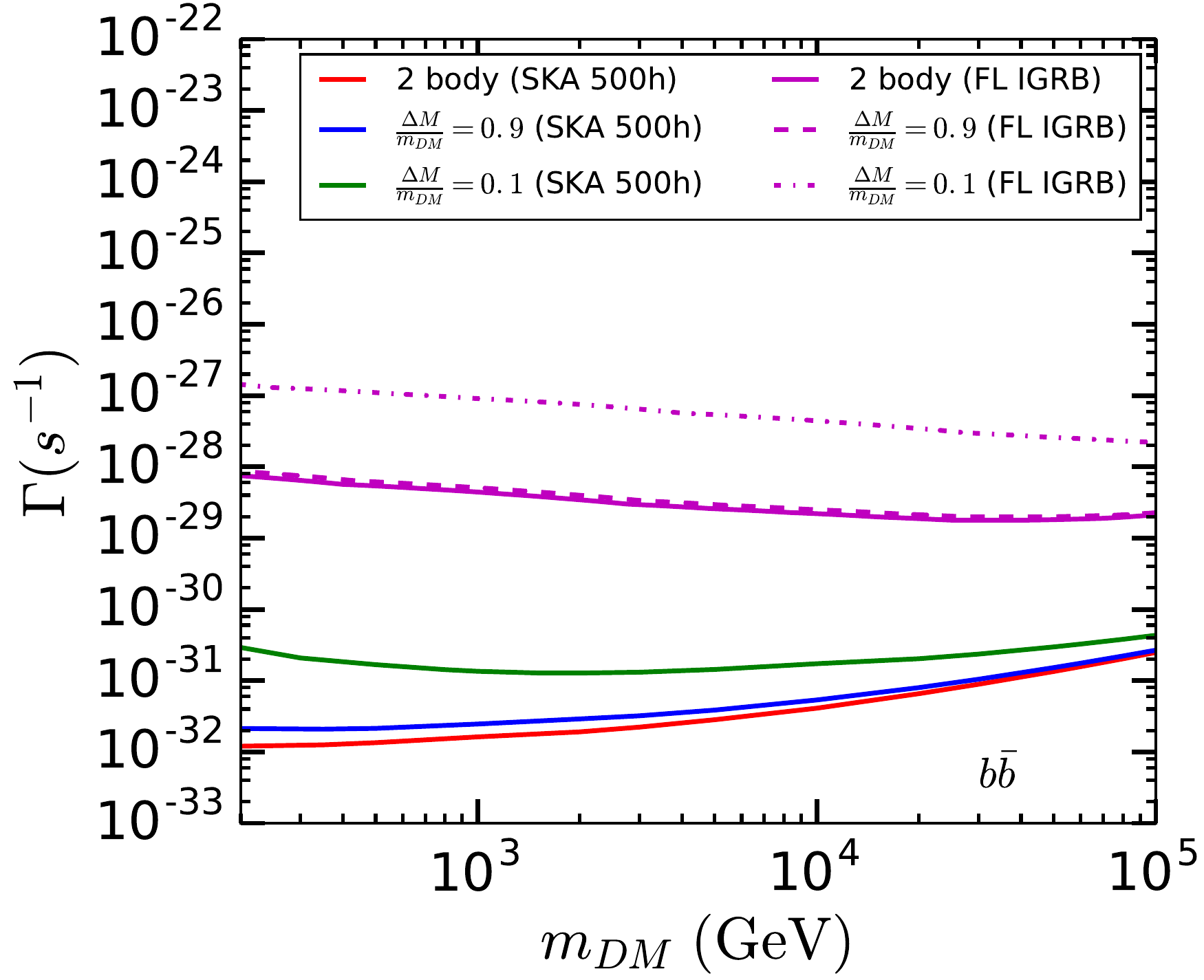}\hspace{2mm}%
  \includegraphics[height=0.4\textwidth, angle=0]{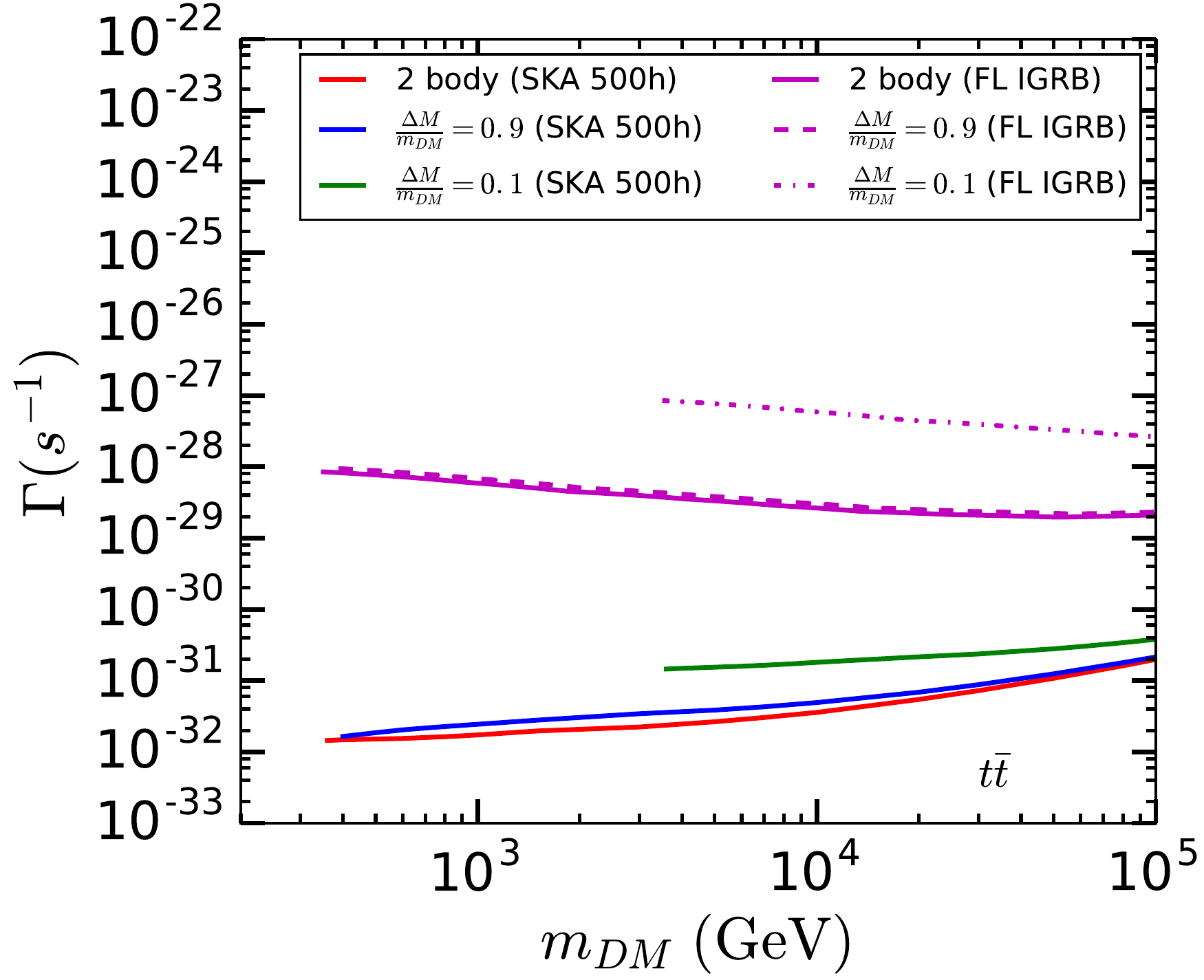}\hspace{2mm}%
  \includegraphics[height=0.4\textwidth, angle=0]{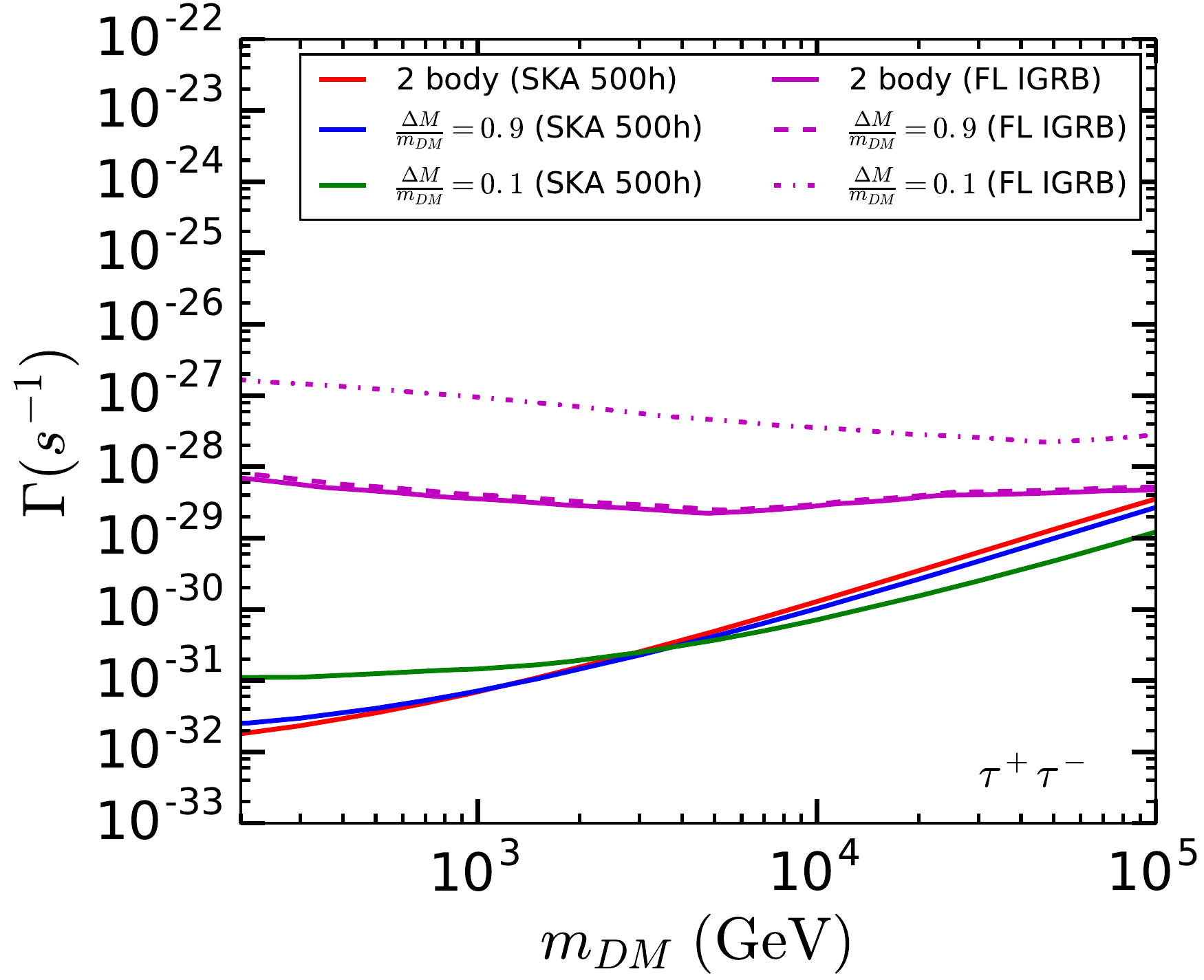}\hspace{2mm}%

  \caption{SKA sensitivity and the upper limit from $\gamma$-ray observation in the $\Gamma - m_{DM}$ plane for the decay of a DM
  (or a dark sector) particle of mass $m_{DM}$ to various fermionic final states ($f\bar{f}$), i.e. either $b\bar{b}$ (upper-left 
  panel) or $t\bar{t}$ (upper-right panel) or $\tau^+\tau^-$ (lower panel). The red, blue and green curves denote the required values 
  of $\Gamma$ to detect the radio fluxes from Draco dSph for the processes $\phi \rightarrow f\bar{f}$ and $\phi_2 \rightarrow \phi_1 f\bar{f}$ 
  (with $\frac{\Delta M}{m_{DM}} = 0.9$ and 0.1), respectively. The (normalised) energy distribution of $f/\bar{f}$ in 
  three-body decay is governed by Eqn. \ref{eqn:ffbar_dist1}. The astrophysical parameters used are $D_0 = 3 \times 10^{28} \mbox{cm}^2 \mbox{s}^{-1}$ 
  and $B = 1$ $\mu G$. The solid, dashed and dashed-dotted magenta lines are the corresponding upper limits on $\Gamma$ from the observation of 
  isotropic $\gamma$-ray background (IGRB) by Fermi-LAT (FL).}
\label{fig:FF_G_mDM}
\end{figure*}

\begin{figure*}[ht!]
\centering
  \includegraphics[height=0.4\textwidth, angle=0]{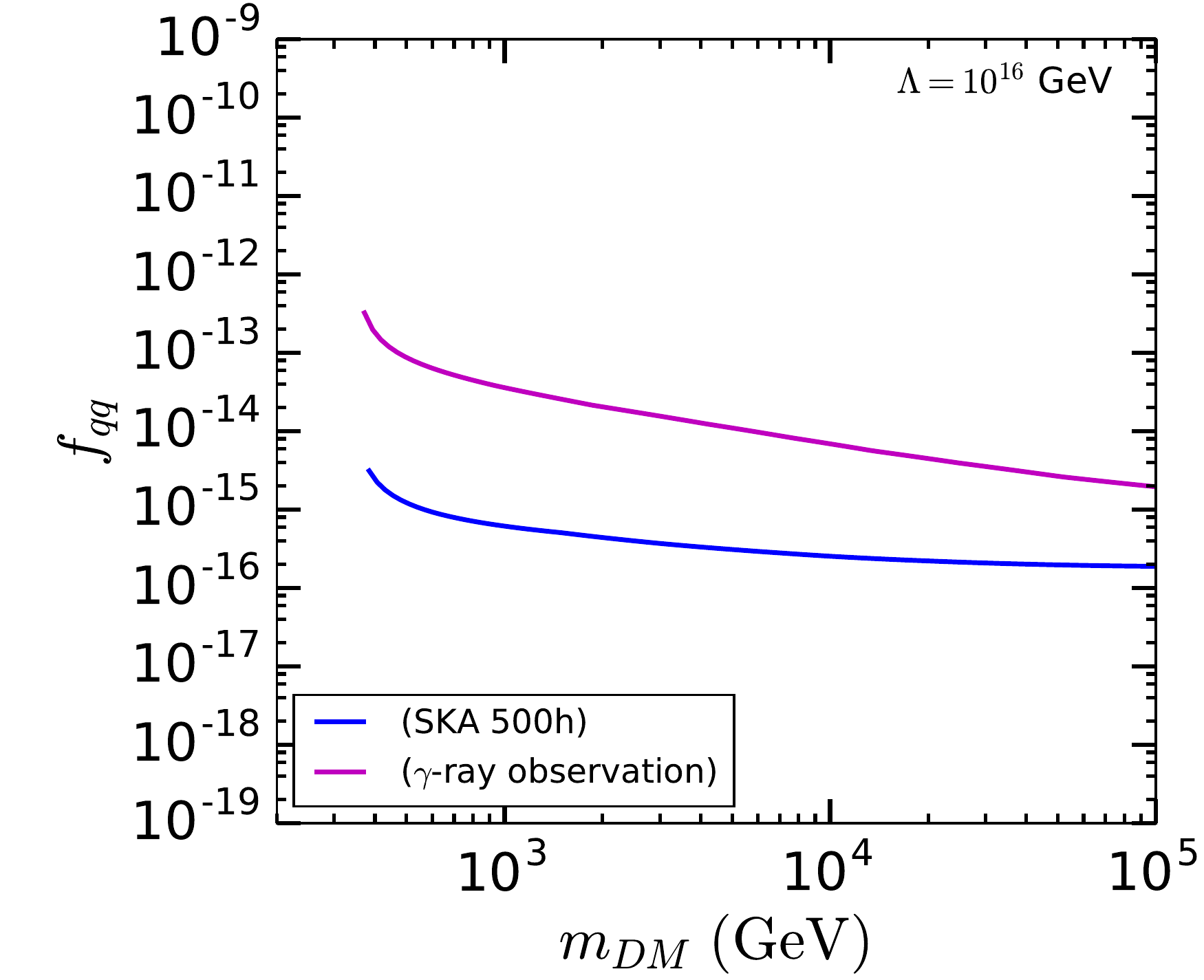}\hspace{2mm}%
  \includegraphics[height=0.4\textwidth, angle=0]{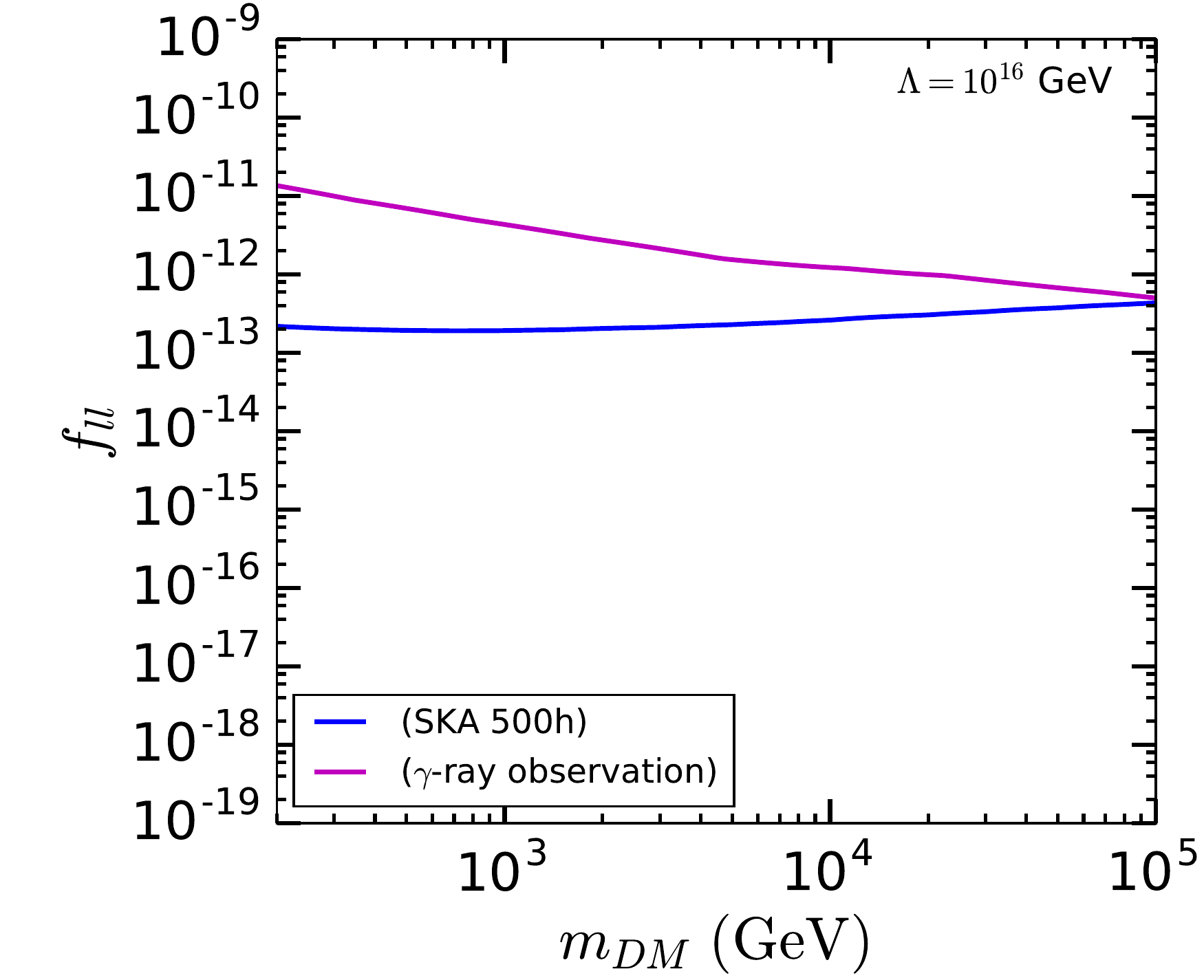}\hspace{2mm}%
  \includegraphics[height=0.4\textwidth, angle=0]{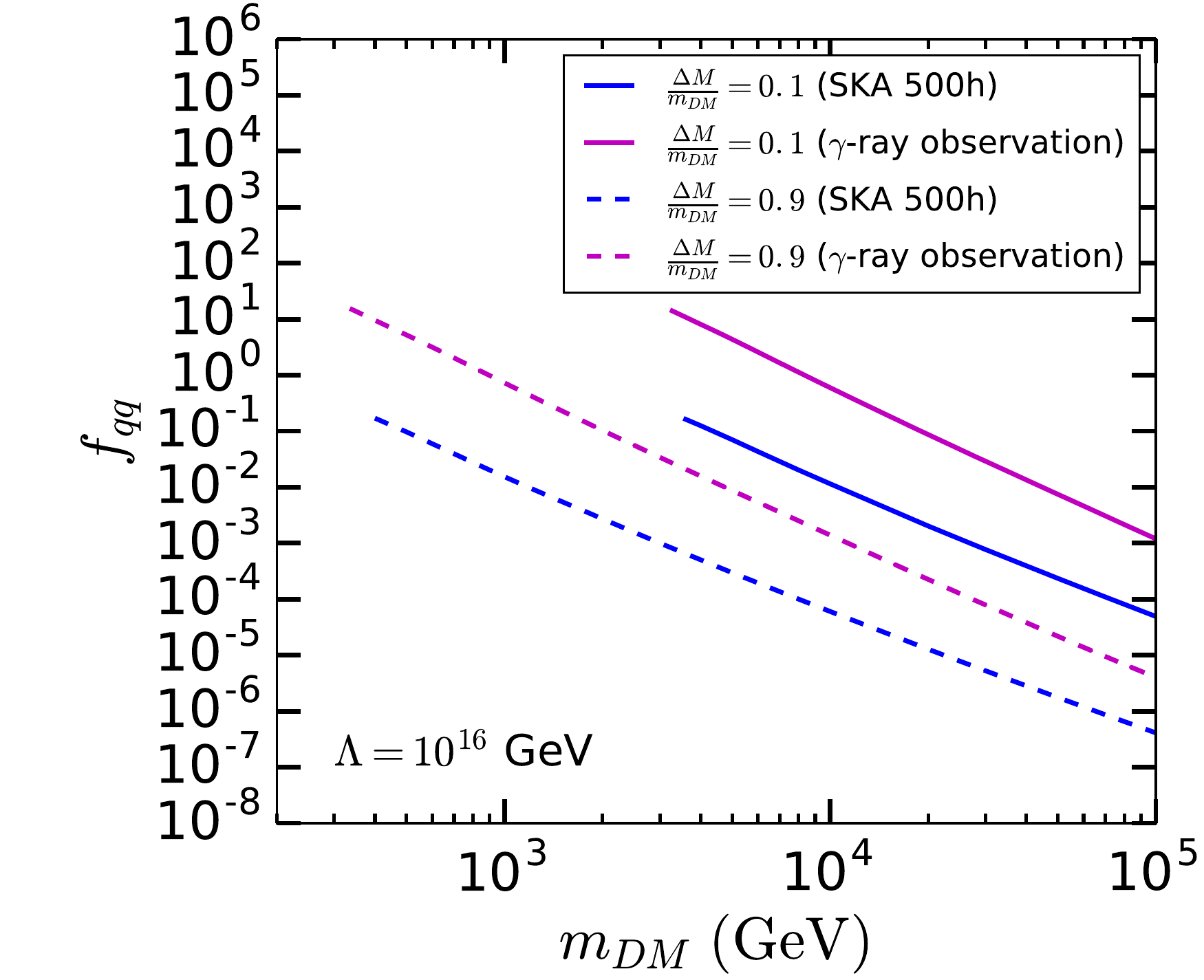}\hspace{2mm}%
  \includegraphics[height=0.4\textwidth, angle=0]{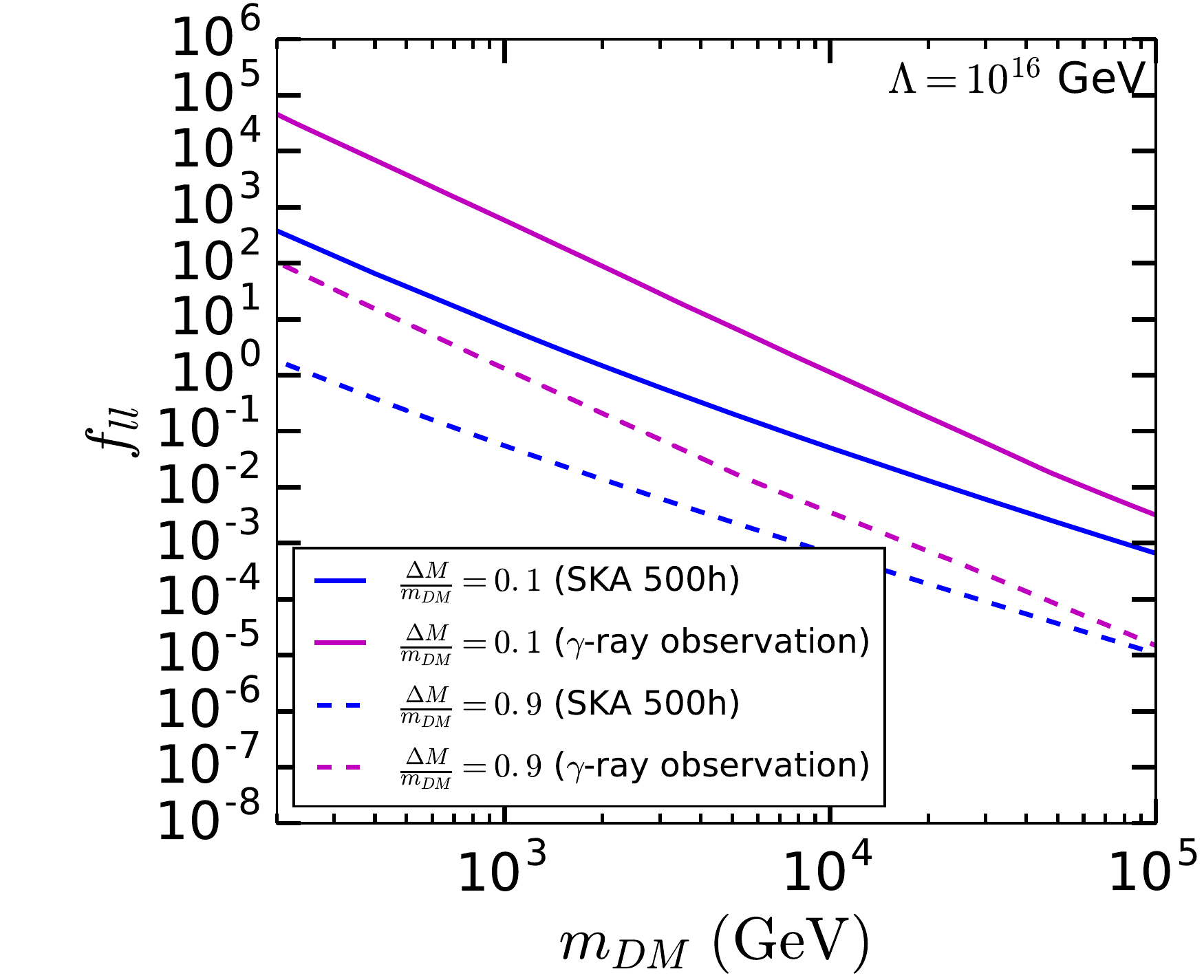}

\caption{{\it Upper panel:} Sensitivity of SKA (blue lines; assuming a 500 hours observation for Draco dSph) to the effective couplings 
of $\mathcal{L}_{dim-5}^{\rm fermion,1}$, shown in Eqn. \ref{eqn:dim5_single}, which give the two-body decays of a DM particle of mass 
$m_{DM}$ to the pairs of fermions. $f_{qq}$ is the coupling to both $b$ and $t$ quarks and $f_{ll}$ is the coupling to $\tau$ lepton.
The astrophysical parameters used are $D_0 = 3 \times 10^{28} \mbox{cm}^2 \mbox{s}^{-1}$ and $B = 1$ $\mu G$. 
The magenta lines are the corresponding limits obtained from the isotropic $\gamma$-ray background (IGRB) observation.
{\it Lower panel:} Similar constraints on the effective couplings of $\mathcal{L}_{dim-6}^{\rm fermion,1}$, shown in Eqn. \ref{eqn:dim6_multi}, 
which give the three-body decays of a DM particle of mass $m_{DM}$. Two different values of $\Delta M /m_{DM}$ have been considered, 
$\Delta M /m_{DM} = 0.1$ (solid lines) and 0.9 (dashed lines).}
\label{fig:fFF_mDM}
\end{figure*}

\begin{figure*}[ht!]
\centering
  \includegraphics[height=0.4\textwidth, angle=0]{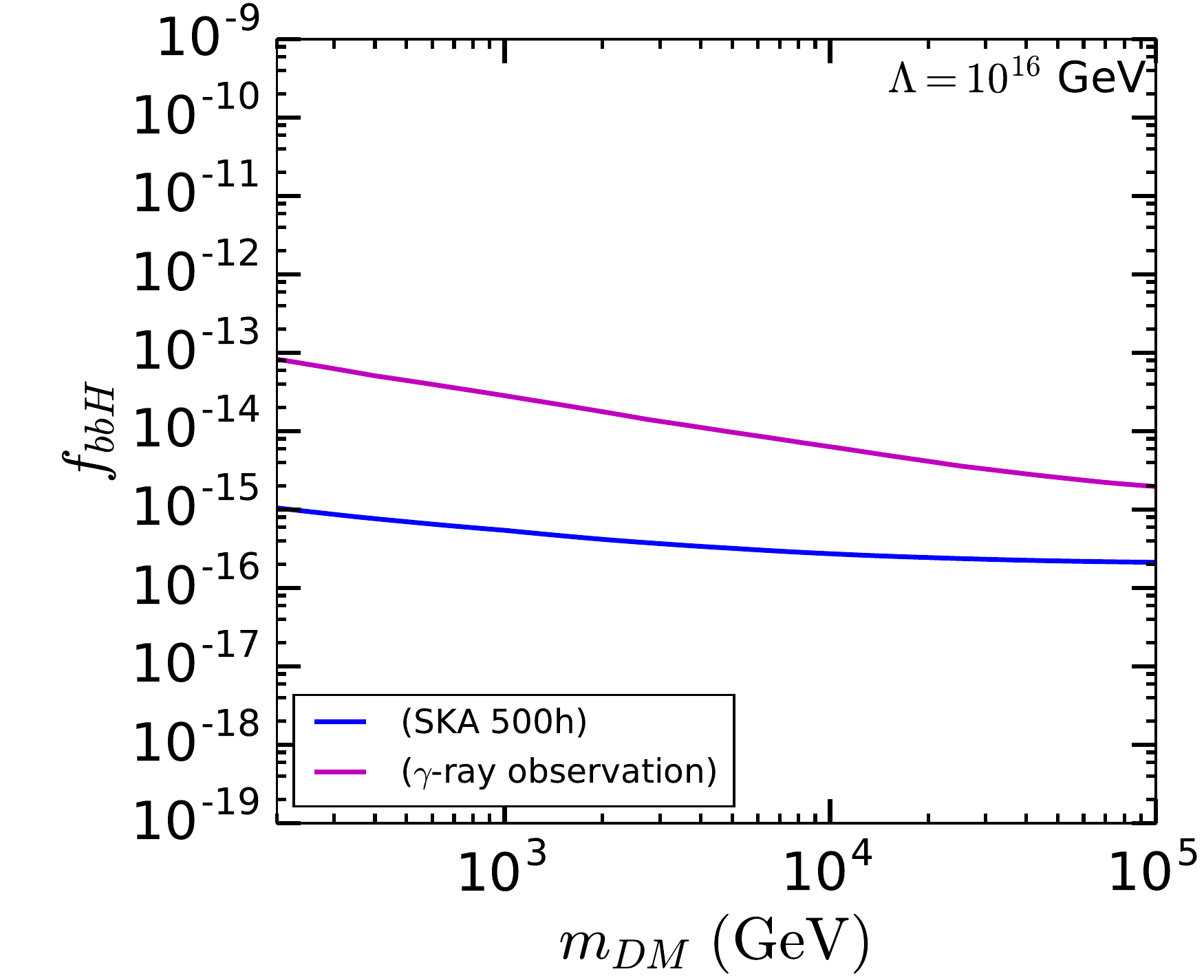}\hspace{2mm}%
  \includegraphics[height=0.4\textwidth, angle=0]{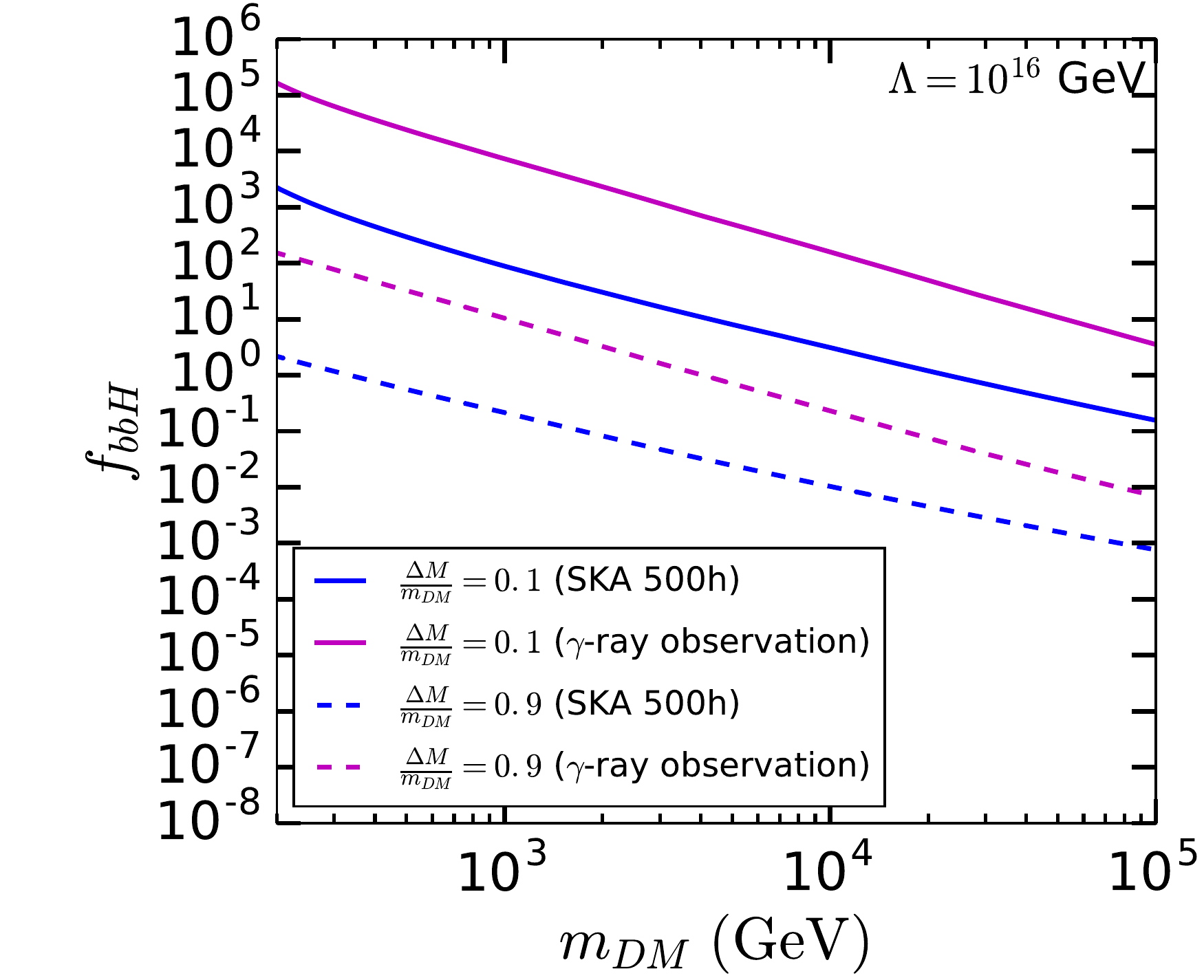}\hspace{2mm}%
  \includegraphics[height=0.4\textwidth, angle=0]{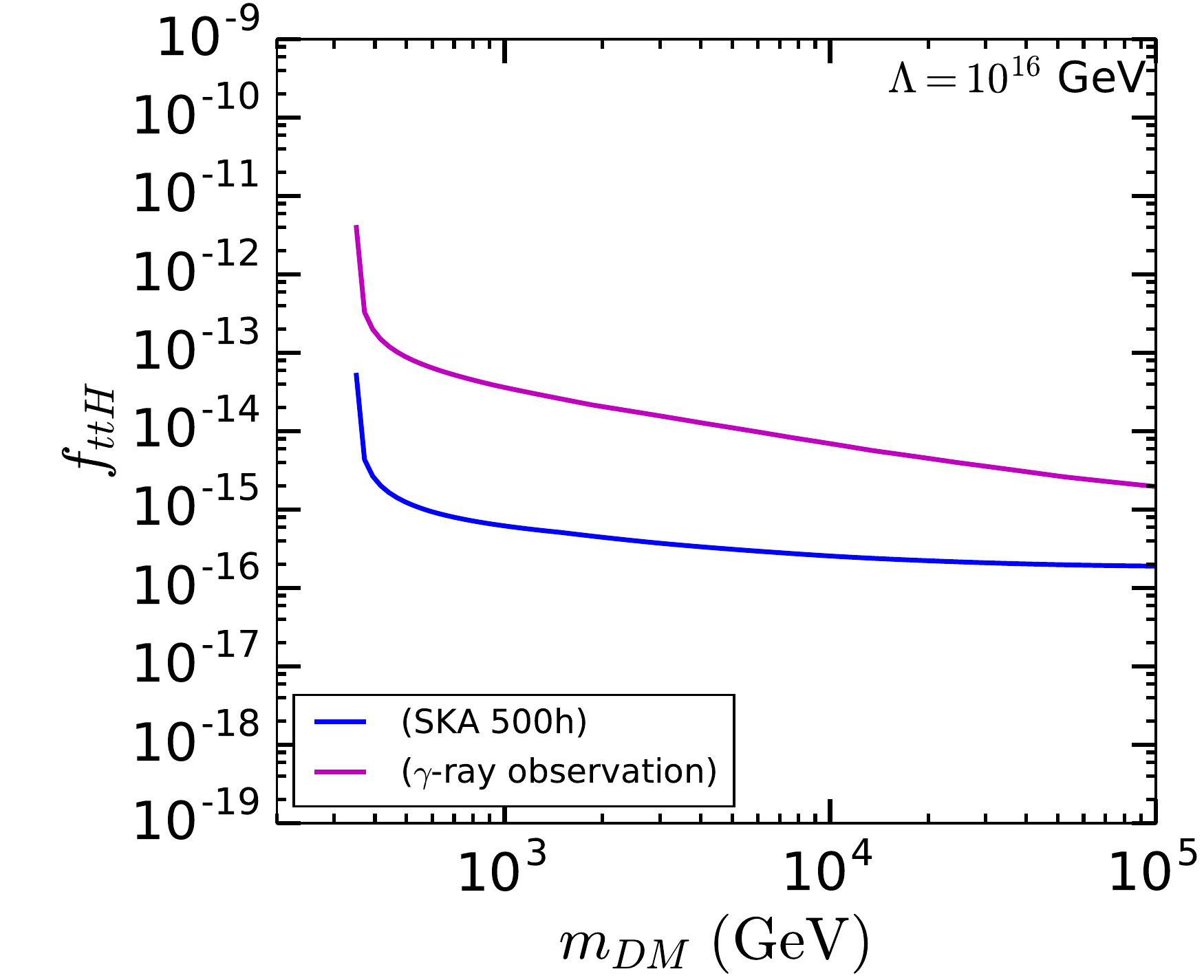}\hspace{2mm}%
  \includegraphics[height=0.4\textwidth, angle=0]{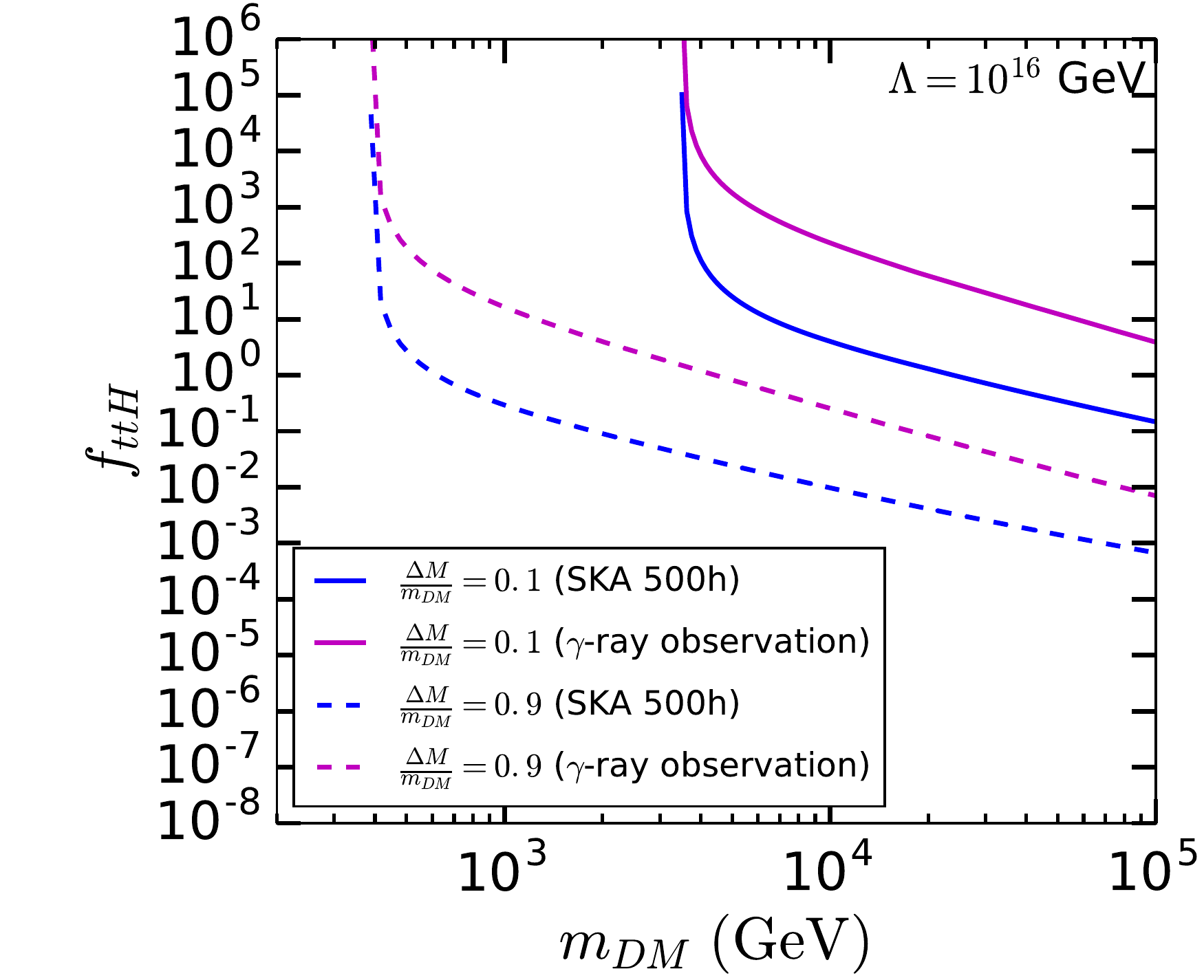}\hspace{2mm}%
  \includegraphics[height=0.4\textwidth, angle=0]{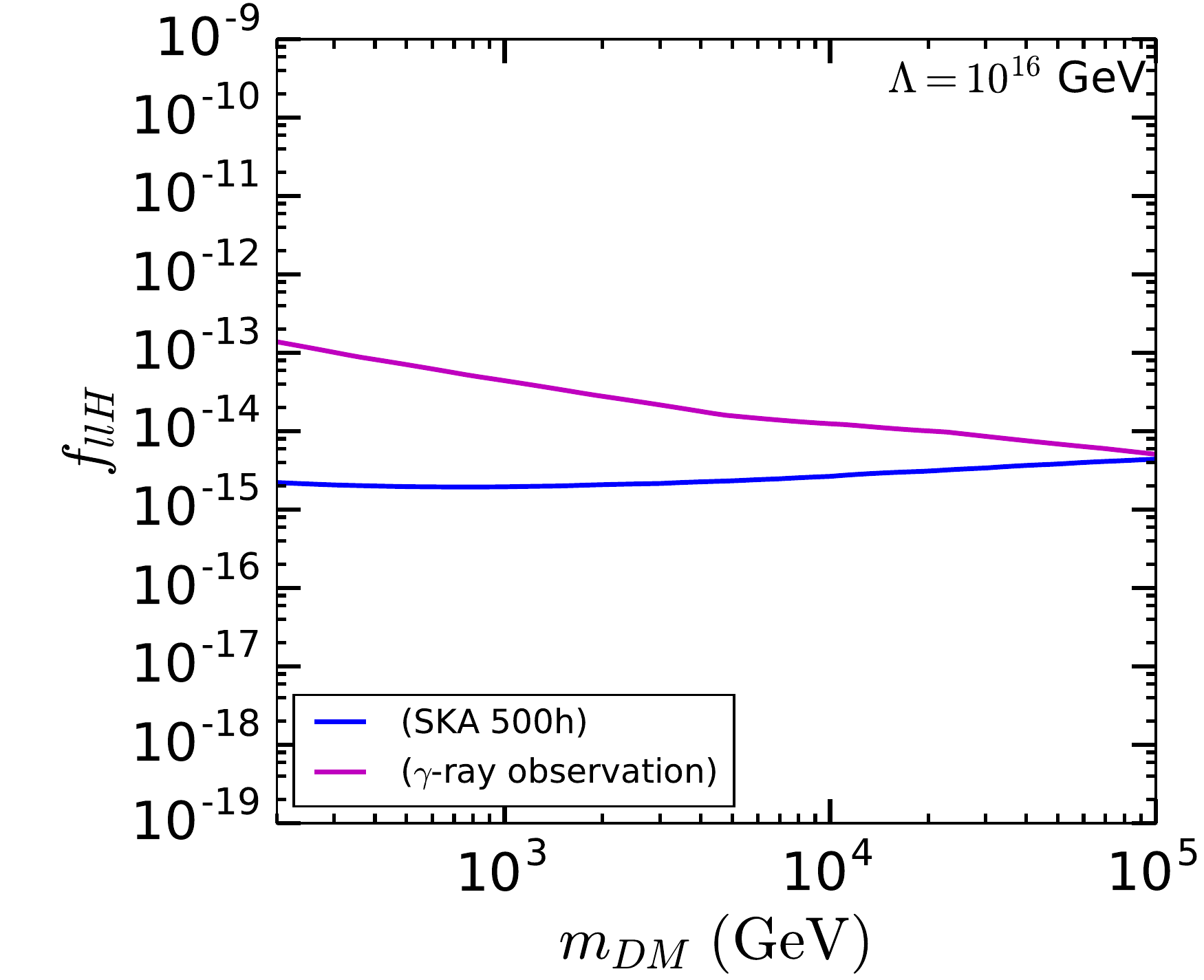}\hspace{2mm}%
  \includegraphics[height=0.4\textwidth, angle=0]{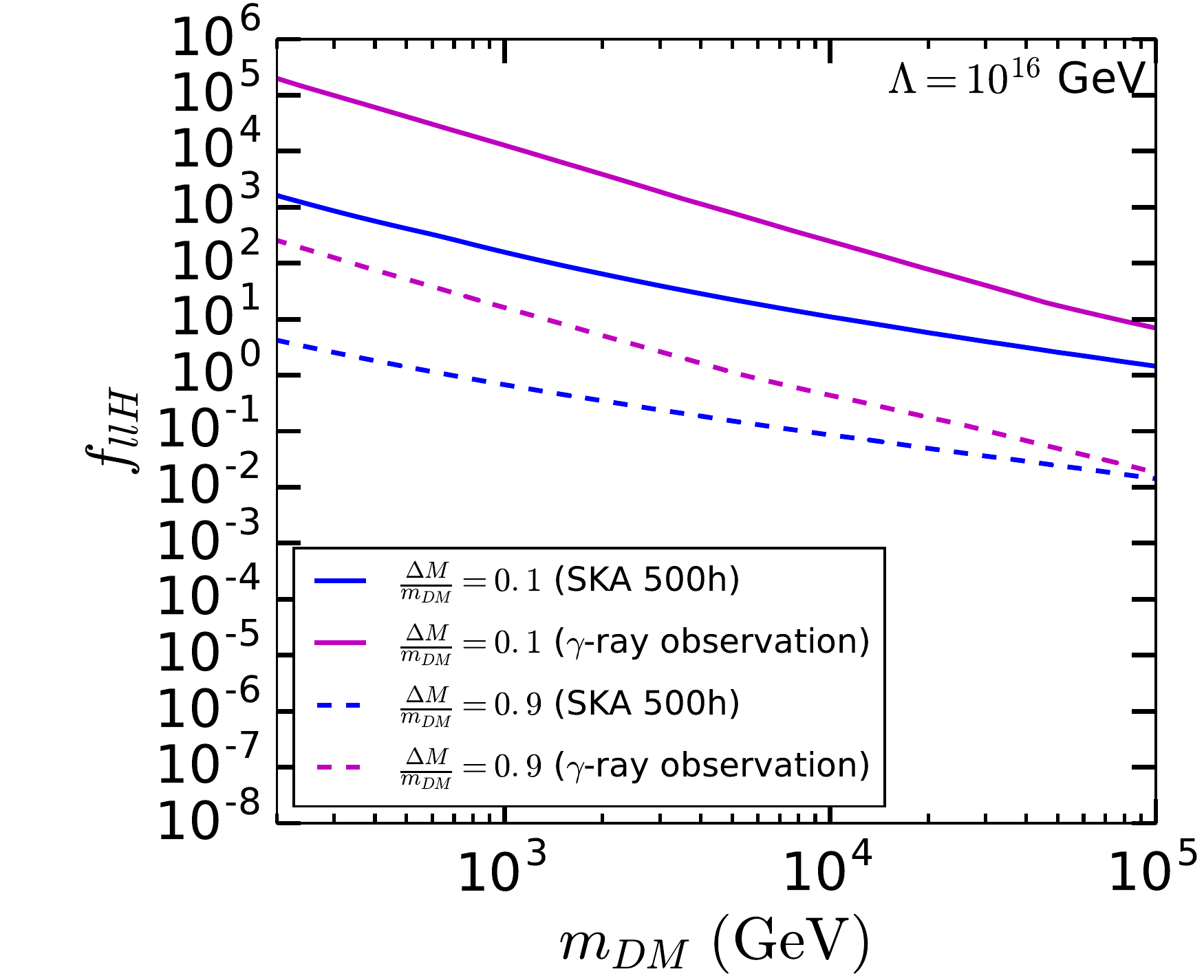}\hspace{2mm}%
  \caption{{\it Left column:} Sensitivity of SKA (blue lines; assuming a 500 hours observation for Draco dSph) to the effective couplings 
  of $\mathcal{L}_{dim-5}^{\rm fermion,2}$, shown in Eqn. \ref{eqn:dim5_single}, which give rise to the two-body decays of a DM particle 
  of mass $m_{DM}$ to a pair of fermions. $f_{bbH}$, $f_{ttH}$ and $f_{llH}$ are the couplings to $b$ quark, $t$ quark and $\tau$ lepton, 
  respectively. The astrophysical parameters used are $D_0 = 3 \times 10^{28} \mbox{cm}^2 \mbox{s}^{-1}$ and $B = 1$ $\mu G$.
  The magenta lines are the corresponding limits obtained from the isotropic $\gamma$-ray background (IGRB) observation.
   {\it Right column:} Similar constraints on the effective couplings of $\mathcal{L}_{dim-6}^{\rm fermion,2}$, shown in 
   Eqn. \ref{eqn:dim6_multi}, which causes the three-body decays of a DM particle of mass $m_{DM}$. Two different values of 
   $\Delta M /m_{DM}$ have been considered, $\Delta M /m_{DM} = 0.1$ (solid lines) and 0.9 (dashed lines).}
\label{fig:fFFH_mDM}
\end{figure*}

\begin{figure*}[ht!]
\centering
  \includegraphics[height=0.4\textwidth, angle=0]{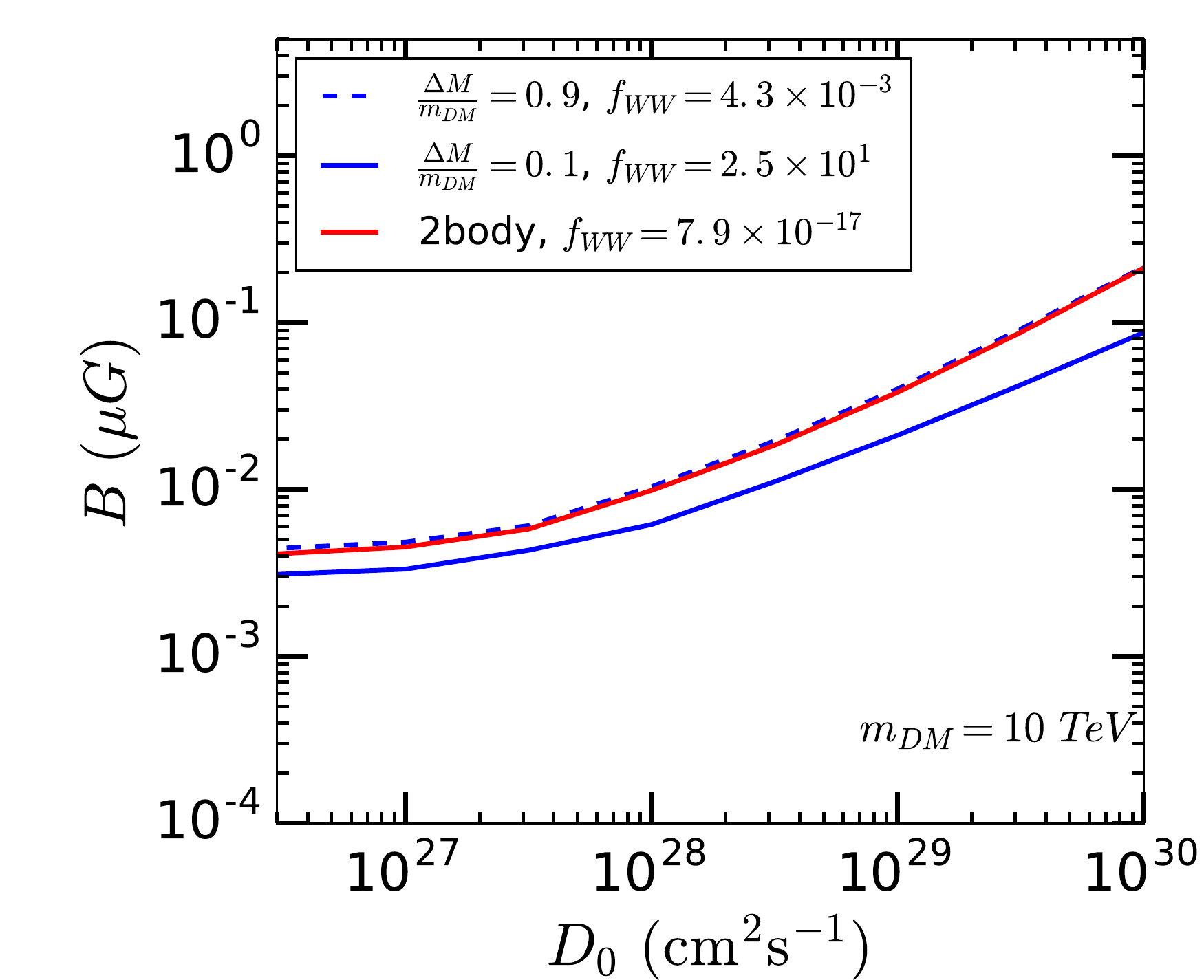}\hspace{2mm}%
  \includegraphics[height=0.4\textwidth, angle=0]{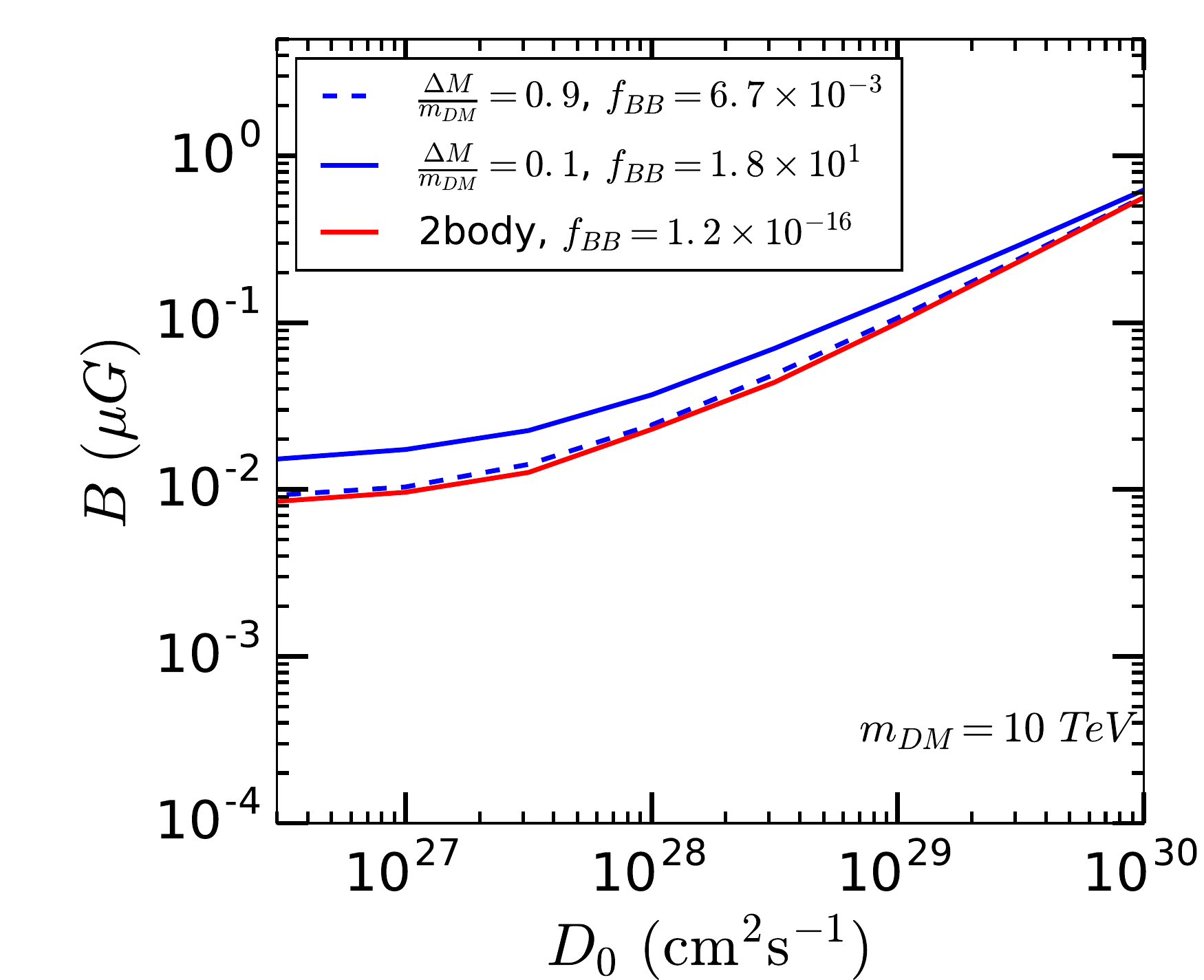}\hspace{2mm}%
  \caption{Limits in the $B - D_0$ plane to observe a DM decay induced radio signal at SKA (500 hours) from Draco dSph. It is assumed that the 
  DM particle is of mass 10 TeV and it decays to various gauge boson final states through the gauge boson effective operators listed in 
  Eqns. \ref{eqn:dim5_single} and \ref{eqn:dim6_multi}. The values of the effective couplings are kept fixed at the upper limits obtained 
  from the isotropic $\gamma$-ray background (IGRB) at this DM mass (see Fig.\ref{fig:fVV_mDM}).}
\label{fig:D0_B_fVV}
\end{figure*}

\begin{figure*}[ht!]
\centering
  \includegraphics[height=0.4\textwidth, angle=0]{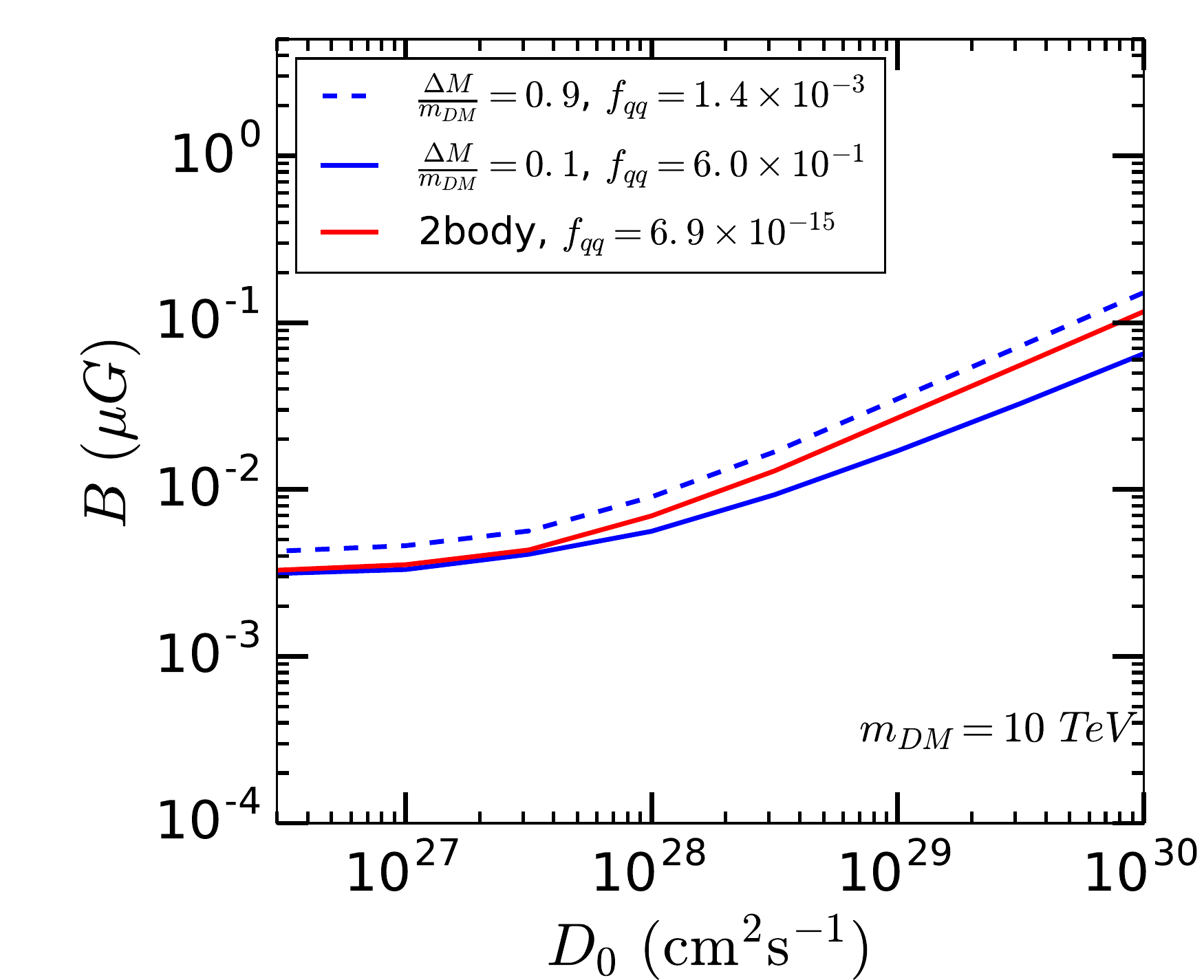}\hspace{2mm}%
  \includegraphics[height=0.4\textwidth, angle=0]{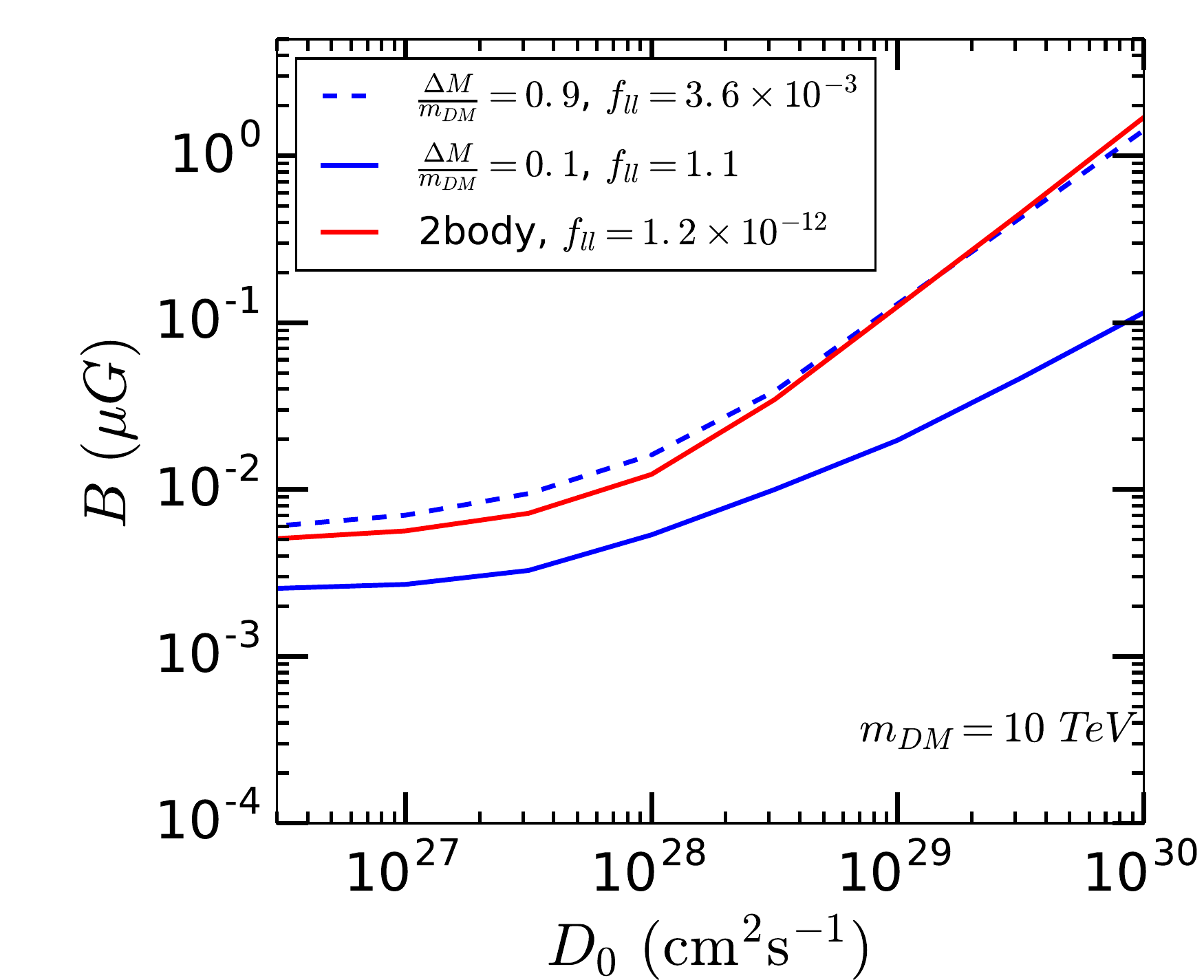}\hspace{2mm}%
  \caption{Limits in the $B - D_0$ plane to observe a DM decay induced radio signal at SKA (500 hours) from Draco dSph. 
  It is assumed that the DM particle is of mass 10 TeV and it decays to various 
fermionic final states through the fermionic operators $\mathcal{L}^{\rm fermion,1}_{dim-5}$ and 
$\mathcal{L}^{\rm fermion,1}_{dim-6}$, listed in Eqns. \ref{eqn:dim5_single} and \ref{eqn:dim6_multi}, respectively. The values of the effective couplings are kept 
fixed at the upper limits obtained from the isotropic $\gamma$-ray background (IGRB) at this DM mass (see Fig.\ref{fig:fFF_mDM}).}
\label{fig:D0_B_fffbarderiv}
\end{figure*}

\begin{figure*}[ht!]
\centering
  \includegraphics[height=0.4\textwidth, angle=0]{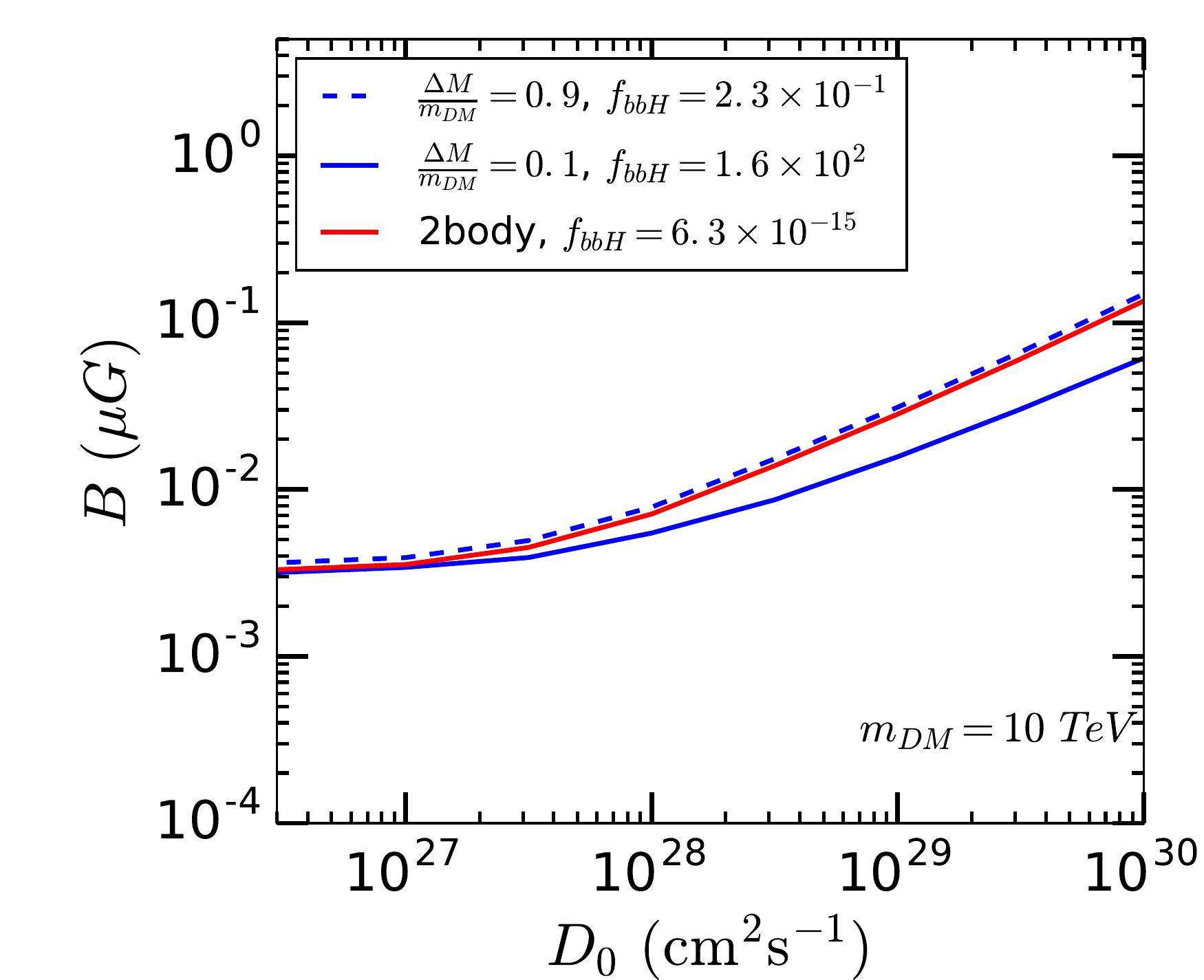}\hspace{2mm}%
  \includegraphics[height=0.4\textwidth, angle=0]{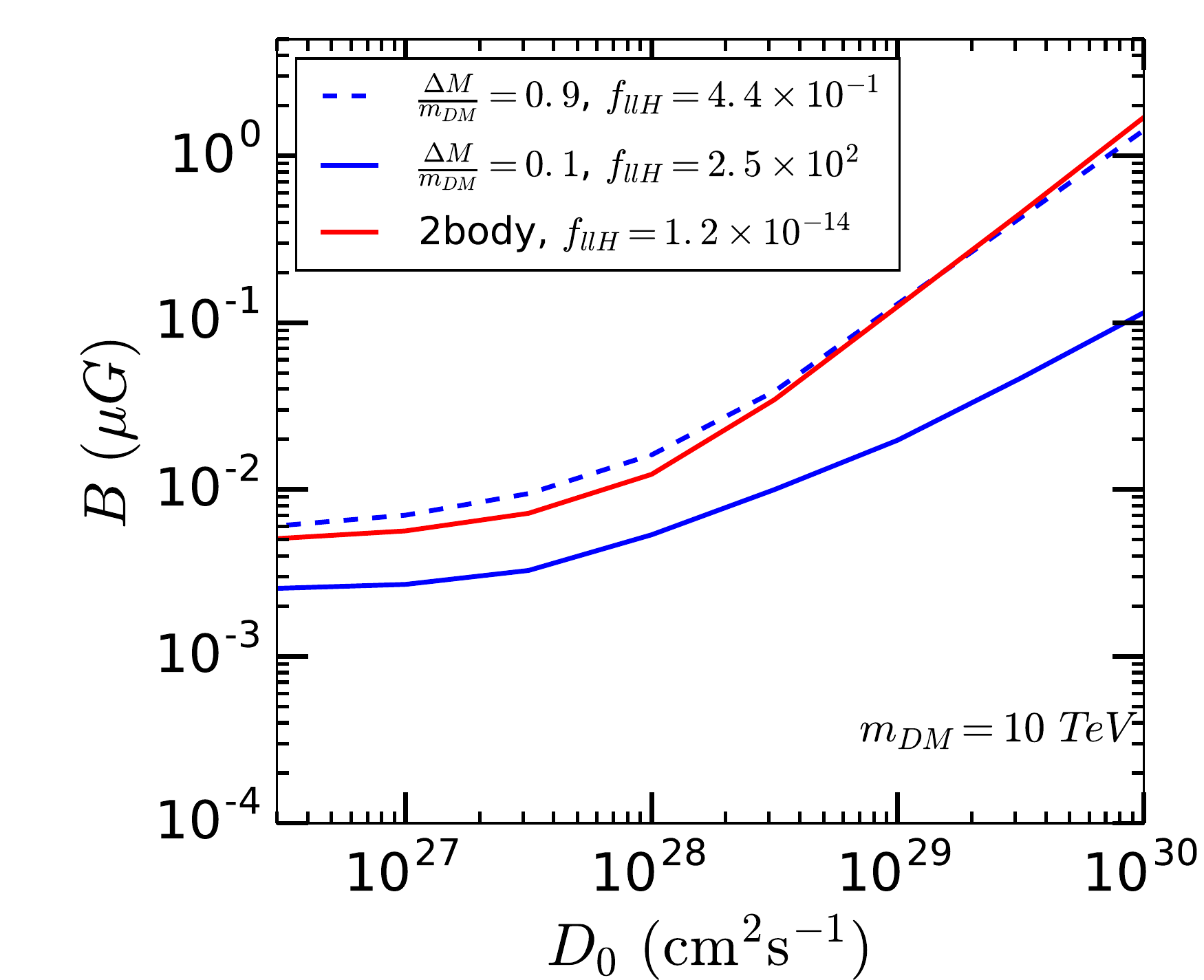}\hspace{2mm}%
  \caption{Limits in the $B - D_0$ plane to observe a DM decay induced radio signal at SKA (500 hours) from Draco dSph. It is assumed that 
  the DM particle is of mass 10 TeV and it decays to various 
fermionic final states through the fermionic operators $\mathcal{L}^{\rm fermion,2}_{dim-5}$ and 
$\mathcal{L}^{\rm fermion,2}_{dim-6}$, listed in Eqns. \ref{eqn:dim5_single} and \ref{eqn:dim6_multi}, respectively. 
The values of the effective couplings are kept fixed at the upper limits obtained from the isotropic $\gamma$-ray background (IGRB) 
at this DM mass (see Fig.\ref{fig:fFF_mDM}).}
\label{fig:D0_B_fffbarHderiv}
\end{figure*}


\end{document}